\documentclass[preprint2]{emulateapj}

\newcommand{\bzcat}{ROMA-BZCAT}

\newcommand{\chn}{{\it Chandra}}

\newcommand{\fer}{{\it Fermi}}

\newcommand{\swf}{{\it Swift}}

\newcommand{\wse}{{\it WISE}}

\usepackage{graphicx}
\usepackage{longtable}
\slugcomment{version \today: fm}
\shorttitle{Unidentified Gamma-ray Sources III}
\shortauthors{F. Massaro et al. 2013}

\begin{document}
\title{Unveiling the nature of the unidentified gamma-ray sources III: \\ gamma-ray blazar-like counterparts at low radio frequencies}
\author{
F. Massaro\altaffilmark{1}, 
R. D'Abrusco\altaffilmark{2}, 
M. Giroletti\altaffilmark{3},
A. Paggi\altaffilmark{2}, 
N. Masetti\altaffilmark{4},
G. Tosti\altaffilmark{5,6},
M. Nori\altaffilmark{7}
\& 
S. Funk\altaffilmark{1}.
}

\altaffiltext{1}{SLAC National Laboratory and Kavli Institute for Particle Astrophysics and Cosmology, 2575 Sand Hill Road, Menlo Park, CA 94025, USA.}
\altaffiltext{2}{Harvard - Smithsonian Astrophysical Observatory, 60 Garden Street, Cambridge, MA 02138, USA.}
\altaffiltext{3}{INAF Istituto di Radioastronomia, via Gobetti 101, 40129, Bologna, Italy.}
\altaffiltext{4}{INAF - Istituto di Astrofisica Spaziale e Fisica Cosmica di Bologna, via Gobetti 101, 40129, Bologna, Italy.}
\altaffiltext{5}{Dipartimento di Fisica, Universit\`a degli Studi di Perugia, 06123 Perugia, Italy.}
\altaffiltext{6}{Istituto Nazionale di Fisica Nucleare, Sezione di Perugia, 06123 Perugia, Italy.}
\altaffiltext{7}{University of Bologna, Department of Physics and Astronomy, viale Berti Pichat 6/2, 40127 Bologna Italy.}

\begin{abstract}
About one third of the $\gamma$-ray sources listed in the 
second \fer\ LAT catalog (2FGL) have no firmly established counterpart at lower energies
so being classified as unidentified gamma-ray sources (UGSs).
Here we propose a new approach to find candidate counterparts for the UGSs
based on the 325 MHz radio survey performed with Westerbork Synthesis Radio Telescope (WSRT)
in the northern hemisphere. First we investigate the low-frequency radio properties of blazars, the largest known 
population of $\gamma$-ray sources; then we search for sources with similar radio properties combining the 
information derived from the Westerbork Northern Sky Survey (WENSS) with those of the 
NRAO VLA Sky survey (NVSS). We present a list of candidate counterparts for 32 UGSs
with at least one counterpart in the WENSS.
We also performed an extensive research in literature to look for infrared and optical counterparts of the $\gamma$-ray
blazar candidates selected with the low-frequency radio observations to confirm their nature.
On the basis of our multifrequency research we identify 23 new $\gamma$-ray blazar candidates
out of 32 UGSs investigated. Comparison with previous results on the UGSs are also presented.
Finally, we speculate on the advantages on the use of the low-frequency radio observations to associate UGSs and to search
for $\gamma$-ray pulsar candidates.
\end{abstract}

\keywords{galaxies: active - galaxies: BL Lacertae objects -  radiation mechanisms: non-thermal}

\section{Introduction}
\label{sec:intro}
Since the epoch of the first $\gamma$-ray surveys performed by COS-B in the 1970s \citep[e.g.,][]{hermsen77} and by the 
Compton Gamma-ray Observatory in the 1990s \citep[e.g.,][]{hartman99}, many different approaches, based on multifrequency 
observations, have been adopted to decrease the number of the unidentified gamma-ray sources (UGSs), 
mostly using radio, optical and X-ray observations \citep[e.g.,][]{thompson08}.
A significant step toward the association of the gamma-ray sources came with the recent lunch of the \fer\ satellite \citep[e.g.,][]{abdo09}.
However, despite the improvements on the source localization provided by \fer,
the large uncertainty regions of the $\gamma$-ray positions makes the 
association of the UGSs still a challenging job.

According to the 2nd \fer\ LAT catalog \citep[2FGL;][]{nolan12}, 
about 1/3 of the $\gamma$-ray detected sources have no assigned counterpart at lower energies. 
A large fraction of the UGSs could be blazars, one of the most peculiar 
class of radio loud active galactic nuclei, being the largest known population shining in the
$\gamma$-ray sky \citep[e.g.,][]{mukherjee97,abdo10}.
However, due to the incompleteness of the current radio and X-ray surveys, it was
not always possible to find the blazar-like counterpart for many of the UGSs.

In the unification scenario blazars fit as radio loud active galaxies \citep[e.g.,][]{blandford78a,blandford79}.
They are compact radio sources with a flat radio spectrum that steepens toward the infrared-optical bands.
Their spectral energy distributions show two main 
broadly peaked components: the low-energy one with its maximum in the IR-to-X-ray frequency range, 
while that at high energies one peaking from MeV to TeV energies.
Their broad band emission features high and variable polarization, 
apparent superluminal motions, and high apparent luminosities, 
coupled with rapid flux variability from the radio to $\gamma$-rays \citep[e.g.,][]{urry95}
and peculiar IR colors \citep{paper1}.

We distinguish between the low luminosity class, constituted by BL Lac objects,
characterized by featureless optical spectra and the flat-spectrum radio quasars 
with optical spectra typical of quasars \citep{stickel91,stoke91,laurent99}.
In the following we label the former class as BZBs and the 
latter as BZQs, following the nomenclature 
of the Multiwavelength Blazar Catalog \citep[\bzcat,][]{massaro09,massaro10a,massaro11}.

Recently, many attempts have been developed to associate and to characterize the UGSs, 
mostly aiming at discovering new $\gamma$-ray blazar candidates not listed in the known catalogs.
Several of these methods were based on the use of pointed \swf\ observations \citep[e.g.,][]{mirabal09a,mirabal09b,paggi13} 
as well as on statistical approaches \citep[e.g.][]{mirabal10,ackermann12}
or on radio follow up observations \citep[e.g.,][]{kovalev09a,kovalev09b,mahony10}. 
Moreover, we addressed the problem of searching $\gamma$-ray blazar candidates as 
counterparts of the UGSs with a new association procedure entirely based
on the Wide-field Infrared Survey Explorer (\wse) all-sky observations \citep{wright10}. 
This association method was motivated by the recent discovery that blazars have distinct infrared (IR) colors 
with respect to other Galactic and extragalactic sources, thus allowing one to find $\gamma$-ray blazar candidates
in the \wse\ all-sky survey \citep[e.g.][]{paper2,paper3,paper4,paper6}.

Here we propose a new approach to the association of the UGSs and the search of $\gamma$-ray blazar
candidates using the low-frequency radio observations performed 
with the Westerbork Synthesis Radio Telescope (WSRT).
We combine the archival observations present in the 
Westerbork Northern Sky Survey \citep[WENSS;][]{rengelink97}
with those of the NRAO Very Large Array Sky survey \citep[NVSS;][]{condon98}
and of the Very Large Array Faint Images of the Radio Sky at Twenty-Centimeters \citep[FIRST;][]{becker95,white97}
to search for $\gamma$-ray blazar candidates.
The WENSS is a low-frequency radio survey that covers the whole sky north of declination $\sim$+28\degr\
at a wavelength of 92 cm (i.e., 325 MHz) to a limiting flux density of approximately 18 mJy at the 5 $\sigma$ level
\citep[see e.g.][for additional details on the survey]{rengelink97}\footnote{http://www.astron.nl/wow/testcode.php?survey=1}.
Radio observations for the WENSS were made in the period 1991-1996 with the WSRT.

Despite a previous investigation at 102 MHz \citep{artyukh81},
the low-frequency radio data is a completely new and unexplored region of the electromagnetic spectrum
for investigating the blazar emission and in particular for associating UGSs, since low frequency radio observations were not 
used either for the $\gamma$-ray source associations in the first and second \fer-LAT catalogs 
\citep[1FGL, 2FGL;][respectively]{abdo10,nolan12} or compiling the
the second Fermi LAT catalog of active galactic nuclei \citep[2LAC;][]{ackermann11a}, or in previous $\gamma$-ray surveys
as the third EGRET catalog \citep{hartman99} or the first AGILE catalog \citep{pittori09}.
As far as we know the WSRT data where only used to search for counterparts of the two UGSs: 
3EG J2016+3657 and 3EG J2021+3716 \citep{mukherjee00}. 

The paper is organized as follows: in Section~\ref{sec:blazar} 
we describe the samples of blazars used in our investigation;
we then search for the counterparts of the blazars listed in the \bzcat\ that lie
in the WENSS footprint, to characterize their low-frequency radio emission,
focusing on those known as $\gamma$-ray emitters.
In Section~\ref{sec:ugs} we search for radio sources that have the same properties 
of the WENSS blazars within the sample of UGSs listed in the 2FGL, and
we also discuss on the IR and optical counterparts of the 
$\gamma$-ray blazar candidates, selected on the basis of their low-frequency radio properties,
to characterize their multifrequency behavior.
Section~\ref{sec:comparison} is devoted to the comparison with previous analyses of the UGSs.
Finally, Section~\ref{sec:summary} is dedicated to our conclusions
while source details are presented in Appendix. 

For our numerical results, we use cgs units unless stated otherwise.
Spectral indices, $\alpha$, are defined by flux density, S$_{\nu}\propto\nu^{-\alpha}$ and
\wse\ magnitudes at the [3.4], [4.6], [12], [22] $\mu$m (i.e., the nominal \wse\ bands)
are in the Vega system respectively.
We use cgs units unless stated otherwise
and we assume a flat cosmology with $H_{\rm 0}=72$ km s$^{-1}$ Mpc$^{-1}$,
$\Omega_{\rm M}=0.26$ and $\Omega_{\Lambda}=0.74$ \citep{dunkley09}.
The most frequent acronyms are listed in Table~\ref{tab:acronym}.
\begin{table}
\caption{List of most frequent used acronyms.}
\begin{tabular}{|lc|}
\hline
Name & Acronym \\
\hline
\noalign{\smallskip}
Multiwavelenght Catalog of blazars & \bzcat\ \\ 
First \fer\ Large Area Telescope catalog & 1FGL \\
Second \fer\ Large Area Telescope catalog & 2FGL \\
Second \fer\ LAT Catalog of AGNs & 2LAC \\
\hline
\noalign{\smallskip}
BL Lac object & BZB \\
Flat Spectrum Radio Quasar & BZQ \\
Blazar of Uncertain type & BZU \\
\hline
\noalign{\smallskip}
Wide-field Infrared Survey Explorer & \wse\ \\
Westerbork Northern Sky Survey & WENSS \\ 
Low radio frequency Blazar sample & LB \\
Low radio frequency Gamma-ray Blazar sample & LGB \\
\noalign{\smallskip}
\hline
\end{tabular}\\
\label{tab:acronym}
\end{table}

\section{Blazars at low radio frequencies}
\label{sec:blazar}
The starting sample used in our analysis is the one presented in the \bzcat\ v4.1, released in August 2012, 
that constitute the most comprehensive catalog of blazars existing in literature listing 3149 sources
\citep[e.g.][]{massaro11}\footnote{www.asdc.asi.it/bzcat/}.
The catalog includes 1220 BZBs, divided as 950 BL Lacs and 270 BL Lac candidates, as defined by \bzcat, 
1707 BZQs and 222  blazars of uncertain type (BZUs) \citep{massaro11}.

The \bzcat\ catalog is not flux limited and it was mainly compiled on the basis of radio, optical and X-ray surveys. 
The large majority of the blazars in the ROMA-BZCAT
lie at high Galactic latitudes, $b$, with only $\sim$6\% at $|b|<$ 15 deg.
It was built on the basis of the following selection criteria, requiring for each source: 
1) a clear radio detection in one of the major radio surveys: the NVSS, \citep{condon98}, FIRST \citep{becker95,white97} 
and the SUMSS \citep{mauch03}, down to mJy flux densities;
2) a compact radio morphology, or (when extended) with one dominant core and a one-sided jet;
3) an optical identification and knowledge of the optical spectrum, needed to establish their class BZBs or BZQs;
4) an isotropic X-ray luminosity larger than $\sim$10$^{43}$ erg s$^{-1}$ when a redshift measurement is available;
5) a radio spectral index measured between 1.4 GHz (or 0.843 GHz) and 5GHz smaller than 0.5 for the BZQs only.
 
All the details about the \bzcat\ catalog and the surveys used to build it 
can be found in Massaro et al. (2009, 2010, 2011b).
We remark that the coordinates reported in the \bzcat\ are not uniform: 
their accuracy is generally { less than} $<$1\arcsec\ { but it could} reach $\sim$ 5\arcsec,
corresponding to the maximum uncertainty on the radio positions of the NVSS, 
for few sources with radio fluxes close to the survey limit \citep{condon98}.

To search for low frequency radio counterparts of the blazars listed in the \bzcat\ 
we used the WENSS catalog available on the HEASARC 
website\footnote{heasarc.gsfc.nasa.gov/W3Browse/radio-catalog/wenss.html},
{ which is the} union of two separate catalogs obtained from the WENSS Website: 
the WENSS Polar Catalog (18186 sources above $\sim$+72\degr declination) 
and the WENSS Main Catalog (211234 sources in the declination region from $\sim$+28\degr\ to $\sim$+76\degr)
\citep[][see also http://heasarc.gsfc.nasa.gov/W3Browse/radio-catalog/wenss.html for additional details]{rengelink97}.
{ In addition we also verified the crossmatches between the \bzcat\ and the WENSS database
using the WENSS catalog available on VizieR\footnote{http://vizier.u-strasbg.fr/viz-bin/VizieR?-source=VIII/62}.}

\subsection{Spatial associations}
\label{sec:radius}
To search for the positional coincidences between the blazars in the \bzcat\ and the radio sources 
listed in the WENSS we adopted the following approach.

We consider the subsample of blazars listed in the \bzcat\ that lie in the footprint of the WENSS survey
at declination above $\sim$+28\degr. This subsample is constituted by 1143 blazars, distinguished in 499 BZBs,
544 BZQs and 100 BZUs. 
In particular, 270 of them are known $\gamma$-ray emitting blazars: 154 BZBs, 94 BZQs and 22 BZUs
associated in the 2LAC catalog \citep{ackermann11a}.

\begin{figure}[] 
           \includegraphics[height=9.5cm,width=7.2cm,angle=-90]{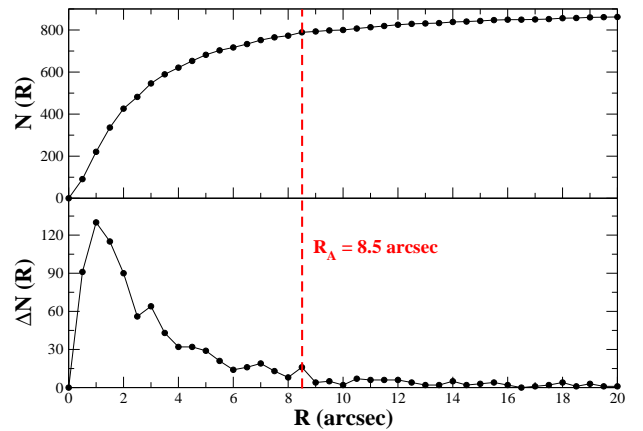}
           \caption{Upper panel) The number of total matches $N(R)$
                        as function of the radius $R$ between 0\arcsec ad 20\arcsec. 
                        Lower panel) The the difference $\Delta N(R)$ between the number of associations at given radius $R$ and those at $R-\Delta R$
                        as function of the radius $R$ in the same range of the above plot.
                        The radial threshold $R_A$ selected for our \bzcat\ - WENSS crossmatches is indicated by
                        the vertical dashed red line (see Section~\ref{sec:radius} for more details).}
          \label{fig:radius}
          \end{figure}

For each of these blazars, we searched for all the WENSS counterparts 
within circular regions of variable radius $R$ in the range between 0\arcsec and 20\arcsec. 
Then, for each value of $R$, we computed the number of correspondences $N(R)$ and
we also calculated the difference between the number of associations at given radius $R$ and those at ($R-\Delta R$),
defined as:
          \begin{equation}
          \Delta N(R) = N(R)-N(R-\Delta R)~,
          \end{equation} 
whereas $\Delta R$= 0.5\arcsec.
Figure~\ref{fig:radius} shows the curves corresponding to $N(R)$, $\Delta N(R)$ for different $R$ values.
For all radii larger than 8\arcsec.5 we found that the increase 
in number of WENSS sources positionally associated with blazars in the \bzcat\
does not vary significantly (i.e., $\Delta N(R)$ sistematically lower than 10), 
this is also highlighted by the flattening of the differential curve of $\Delta N(R)$;
thus we chose the value of $R_A$=8\arcsec.5 as our radial threshold for the counterparts of \bzcat\ blazars in the WENSS.

The number of correspondences between the \bzcat\ and the WENSS is 874 out of the 1143  
(i.e., $\sim$76\%) all unique matches within 8.5\arcsec.
According to the WENSS classification 790 of the are single component sources (flag ``S"),
6 are multiple component sources (flag ``M") while 78 were
labeled as component of a multi-component source (flag ``C") in the WENSS catalog.
In particular, 7 WENSS sources out of the 790 single component ones,
have a fit problem flag and will be excluded from the following analysis \citep[see][for more details on the WENSS catalog flags]{rengelink97}.
The probability of spurious associations is extremely small, being $\sim$ 0.1\% 
\citep[see][and references therein for details on the method to estimate the fraction of spurious associations]{maselli10a,maselli10b,paper1,paper6}.

\subsection{Sample selection}
\label{sec:sample}
We define the following two samples for our investigation of blazars at low radio frequencies.

The first sample, labeled as Low radio frequency Blazar (LB) sample, 
is constituted by all the 789 blazars that have a radio counterpart in the WENSS within 8\arcsec.5
that is a single or a multi-component source (only flags ``S" or ``M").
This LB sample includes 286 BZBs (i.e., 36\%), 429 BZQs (i.e., 54\%) and 74 BZUs (i.e., 10\%).

The second sample, labeled as Low radio frequency Gamma-ray Blazar (LGB) sample,
is a subsample of the LB one, constituted by only the $\gamma$-ray emitting blazars,
so listing 216 sources out the 270 that lie in the footprint of the WENSS.
This LGB sample includes 113 BZBs (i.e., 52\%), 83 BZQs (i.e., 38\%) and 20 BZUs (i.e., 10\%).

It is worth noting that there are more BZBs in the LGB sample than BZQs, opposite to what happens in the associations of the whole LB sample.
This is in agreement with the fact that the $\gamma$-ray detection rate of BZBs is higher than the one of the BZQs, as expected
from previous studies \citep[e.g.,][]{ackermann11a,linford12}.

\begin{figure}[!t] 
           \includegraphics[height=9.5cm,width=7.cm,angle=-90]{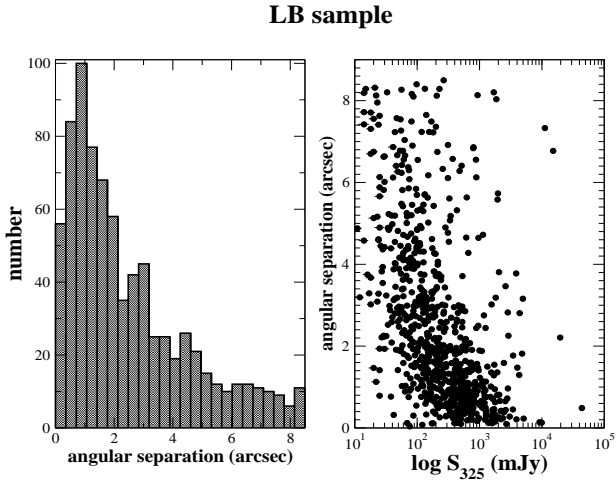}
           \caption{The angular separation between the \bzcat\ positions and that of the WENSS catalog
                         for the whole LB sample (left panel).
                         The scatter plot of \bzcat\ - WENSS angular separation versus the flux density at 325 MHz, $S_{325}$, (right panel).}
          \label{fig:angular}
          \end{figure}

In Figure~\ref{fig:angular} we show the distribution of the angular separation between the \bzcat\ positions and that of the WENSS catalog
for the whole LB sample, together with the scatter plot of the angular separation versus the flux density $S_{325}$ at 325 MHz.
There is a mild trend between the latter two quantities, since, as expected, the position for faint WENSS sources are less accurately determined \citep{rengelink97}. 
In particular, this relation between the flux density and the angular separation
is expected because the positional error of the WENSS survey are scaling as $\approx (\sigma_{rms}/S_{325})$,
where $\sigma_{rms}$ is the local signal to noise \citep[see Section 3.5 of][for more details]{rengelink97}.

\subsection{Radio spectral index}
\label{sec:index}
Since all the sources in the \bzcat\ are already associated with the NVSS, their radio flux densities $S_{1400}$ 
at 1.4 GHz are reported in this catalog. We then define a low frequency radio spectral index: $\alpha_{325}^{1400}$, 
using the $S_{325}$ from the WENSS as:
\begin{equation}
\alpha_{325}^{1400} = - 1.58 \cdot log \left(\frac{S_{1400}}{S_{325}} \right)
\end{equation}
where the factor 1.58 is the $[log(1400/325)]^{-1}$ and both flux densities are measured in units of mJy.
          \begin{figure}[!t] 
           \includegraphics[height=9.5cm,width=7.cm,angle=-90]{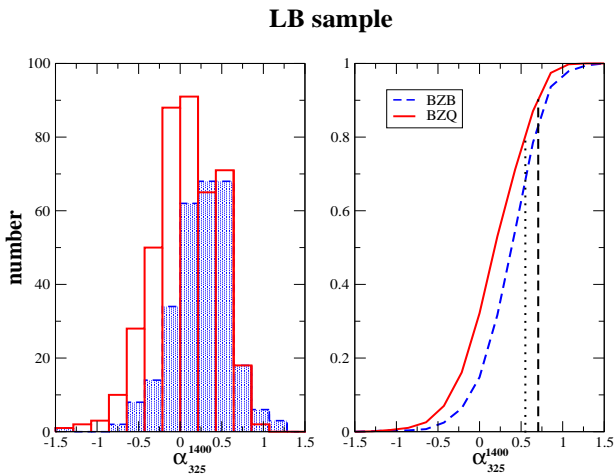}
           \caption{The distributions of the radio spectral index $\alpha_{325}^{1400}$
                       of all the identified blazars in the LB sample, BZBs (blue) and BZQs (red).
                       The two vertical, dashed black, lines in the left panel mark the 0.55 (dotted) and 0.65 (dashed) 
                       values of $\alpha_{325}^{1400}$. Spectral indices for the BZUs are not shown.}
          \label{fig:hist_bzcat}
          \end{figure}

\begin{figure}[!b] 
           \includegraphics[height=9.5cm,width=7.5cm,angle=-90]{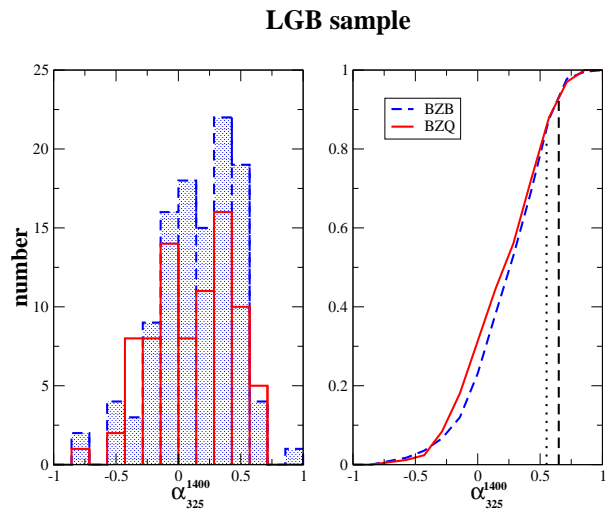}
          \caption{The distributions of the radio spectral index $\alpha_{325}^{1400}$
                       of all the identified blazars in the LGB sample, BZBs (blue) and BZQs (red).
                       It is clear how restricting to the LGB sample the fraction of sources with $\alpha_{325}^{1400}>$1
                       strongly decreases.
                       The two vertical lines mark the 0.55 (dotted) and 0.65 (dashed) 
                       values of $\alpha_{325}^{1400}$. Spectral indices for the BZUs are not shown.}
          \label{fig:hist_fermi}
          \end{figure}

\begin{figure}[] 
           \includegraphics[height=9.5cm,width=7.cm,angle=-90]{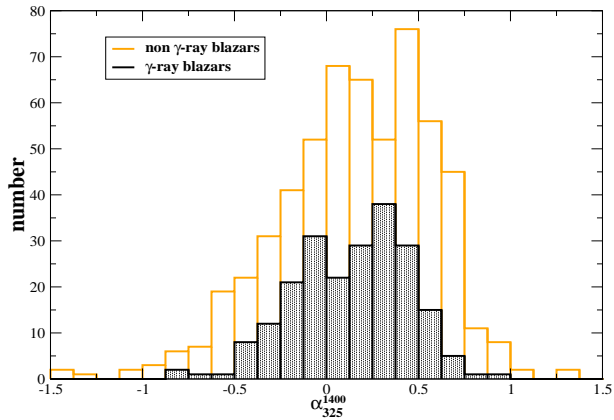}
           \caption{The $\alpha_{325}^{1400}$ distribution of the WENSS blazar
                        non detected by \fer\ in comparison with those with a $\gamma$-ray counterpart.}
          \label{fig:histo1}
          \end{figure}

In Figure \ref{fig:hist_bzcat}, we show the distributions of the $\alpha_{325}^{1400}$
of all the identified blazars in the LB sample distinguishing between the BZBs and BZQs.
The large fraction of radio spectral indices, $\alpha_{325}^{1400}$, are systematically smaller than 1.0 
(i.e., $\sim$99\%), with the 80\% lower than 0.5. 
{ The peak positions of the $\alpha_{325}^{1400}$ distributions in the LB sample (see Figure~\ref{fig:hist_bzcat}) are different, 
since the BZB one peaks at $\alpha_{325}^{1400}$=0.25 while the BZQs at $\alpha_{325}^{1400}$=0.08, however 
performing a Kolmogorov-Smirnov (KS) test, these two distributions are similar at 99\% level of confidence.}
In Figure~\ref{fig:hist_fermi}, the comparison between the BZBs and the BZQs restricting to the LGB sample is also shown.
{ Also in this case the two distributions of BZBs and the BZQs in the LGB sample are also similar at 99\% level of confidence evaluated using the KS test. 
Here we also} note that comparing the distributions reported in Figure \ref{fig:hist_bzcat} with those showed in Figure \ref{fig:hist_fermi},
the fraction of sources with $\alpha_{325}^{1400}>$1 in the LGB sample is smaller than for the entire LB sample.
We also indicate the spectral index values of 0.55 and 0.65, in Figure~\ref{fig:hist_bzcat} and Figure~\ref{fig:hist_fermi}, respectively,
that corresponds to the criteria defined in Section~\ref{sec:assoc} to recognize $\gamma$-ray blazar candidates.
Then, in Figure~\ref{fig:histo1} we present the comparison between 
the distribution of the spectral index $\alpha_{325}^{1400}$ between the $\gamma$-ray emitting blazars 
with those in the WENSS survey non-associated with \fer\ sources. 
{ A KS test indicates that these two distributions are similar at 99\% level of confidence}
   
Flat radio spectra are indeed expected for blazar-like sources when computed considering 
high radio frequency data in the GHz energy range \citep[e.g.,][]{healey07,ivezic02} 
and this spectral property was also used for the identification of $\gamma$-ray sources
since the EGRET era \citep[e.g.,][]{mattox97}.
{ However the low-frequency radio observations, never previously used 
to search for $\gamma$-ray blazar candidates as counterparts of the UGSs, 
clearly show that blazars have flat spectra also around 325 MHz.}

\begin{figure}[!t] 
           \includegraphics[height=9.5cm,width=7.cm,angle=-90]{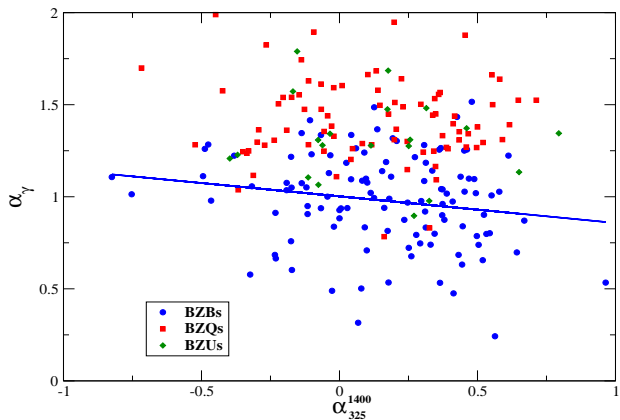}
           \caption{The scatter plot of the radio spectral index $\alpha_{325}^{1400}$ 
                       with respect to that in the $\gamma$-rays $\alpha_{\gamma}$
                       for the BZBs (blue) the BZQs (red) and the BZUs (green) that belong to the LGB sample.
                       No clear trend or correlation was found between the $\alpha_{325}^{1400}$ and $\alpha_{\gamma}$.
                       The main dichotomy between the two classes appear to be highlighted mainly by the $\gamma$-ray spectral shape.
                       The regression line computed for the BZBs only is shown (blue line, see Section~\ref{sec:index} for more details).}
          \label{fig:alphas}
          \end{figure}

Finally, we remark that no correlation or net trend was found between the $\alpha_{325}^{1400}$ and the $\gamma$-ray 
spectral index $\alpha_{\gamma}$ as shown in Figure~\ref{fig:alphas} for the whole LGB sample,
{ while only a marginal trend seems to be present for the BZBs, with the value of the 
Pearson and of the Spearman's rank correlation coefficients of -0.18 and -0.15, respectively.}

\subsection{Radio flux density and luminosity}
\label{sec:flux}
\begin{figure}[] 
           \includegraphics[height=9.5cm,width=7.cm,angle=-90]{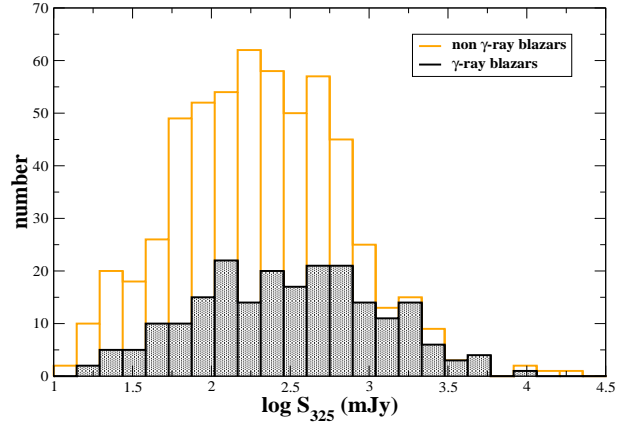}
           \caption{The distribution of the radio flux density $S_{325}$ for the WENSS blazar
                        non detected by \fer\ in comparison with those blazars with a $\gamma$-ray counterpart in the 2FGL.}
          \label{fig:histo2}
          \end{figure}

We computed the distributions of the flux densities for both the considered samples.
We noted that the error on the flux density $S_{325}$ are very small, being 
lower than $\sim$3\% for 68\% of the LB sample and lower than$\sim$2.5\% 
when considering the 68\% of the LGB subsample,
while they are lower than $\sim$7.5\% and lower than $\sim$6.5\% { for} 90\% 
of the sources in the LB and the LGB sample, respectively.
          \begin{figure}[] 
           \includegraphics[height=9.5cm,width=7.cm,angle=-90]{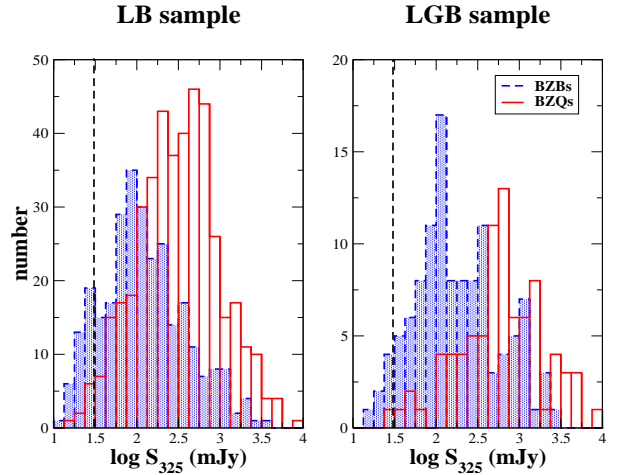}
           \caption{The distributions of the low frequency radio flux density $S_{325}$
                       of all the identified blazars in the LB (left panel) and in the LGB (right panel) 
                       sample, BZBs (blue) and BZQs (red).
                       The dashed vertical line mark the 30 mJy threshold that corresponds 
                       to the limit where the survey is complete.
                       The two distributions appear different { at 90\% level of confidence evaluated with a KS test}
                       in both LB and LGB samples where BZBs are on average fainter than BZQs. 
                       Flux densities for the BZUs are not shown}
          \label{fig:hist_flux}
          \end{figure}
Then, in Figure~\ref{fig:hist_flux}, 
we present the distributions of the low frequency flux density at 325 MHz
{ where it is quite evident that BZBs are, on average, fainter that BZQs in both the LB and LGB samples,
at 90\% level of confidence evaluated with a KS test.}
In Figure~\ref{fig:histo2} the comparison between the distribution of the 
flux densities of the $\gamma$-ray emitting blazars 
with those in the WENSS non-associated with 2FGL sources is also shown.
{ The hypothesis of these two distributions being distinct can be rejected at 99\% significance level.} 

\begin{figure*}[] 
           \includegraphics[height=9.5cm,width=7.cm,angle=-90]{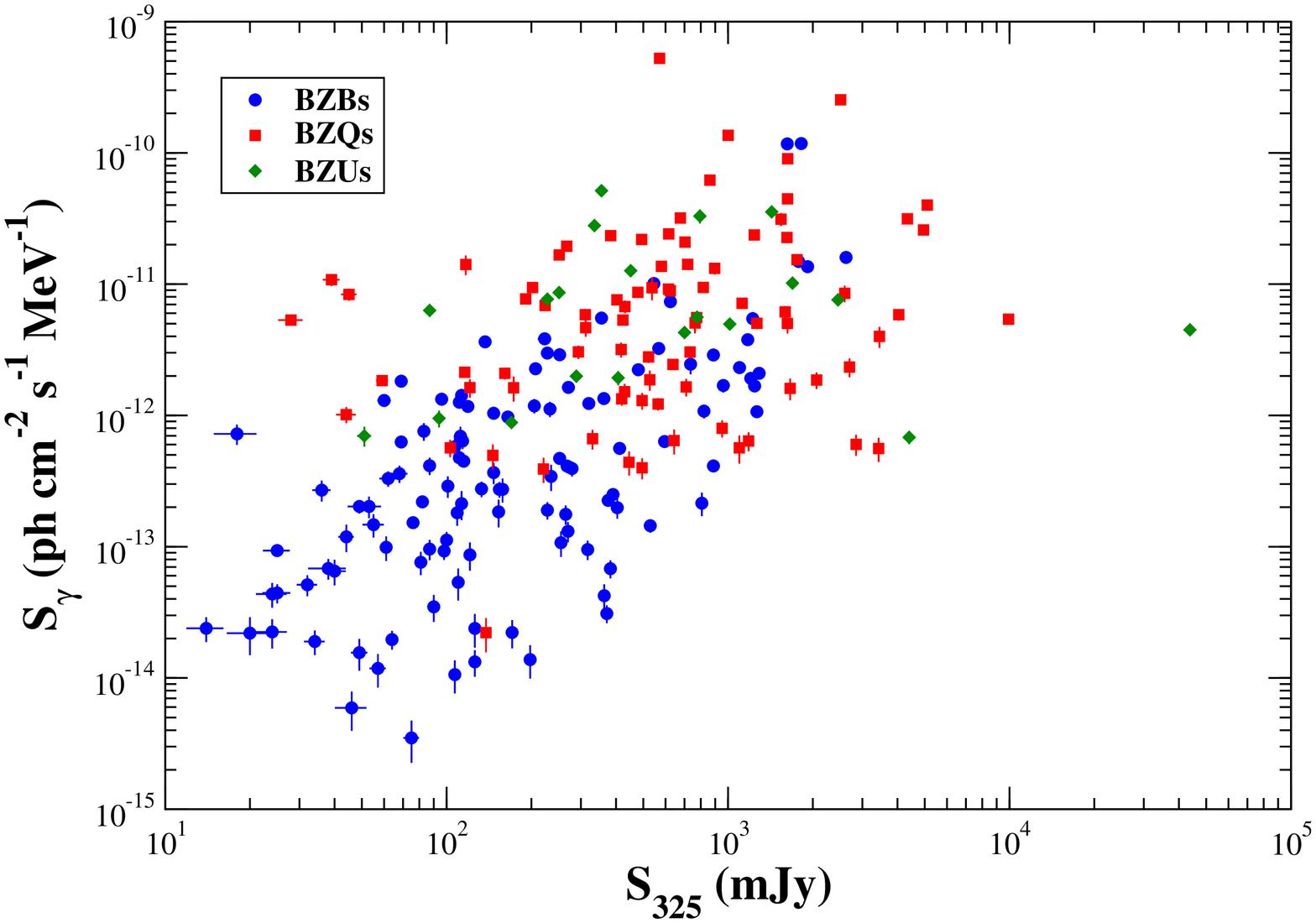}
           \includegraphics[height=9.5cm,width=7.cm,angle=-90]{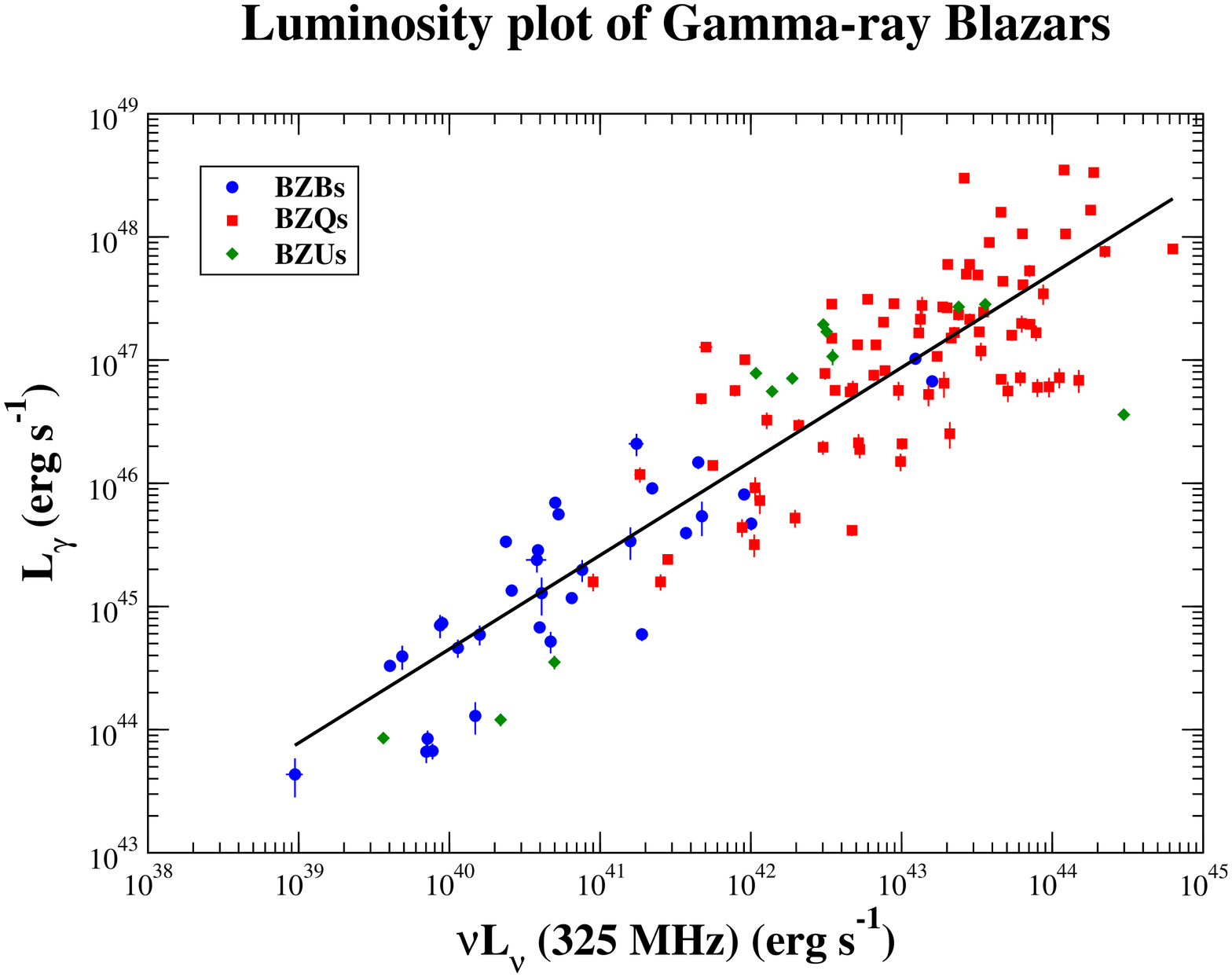}
           \caption{{\it Left panel)} The scatter plot of the radio $vs$ the 
                        $\gamma$-ray flux density for the BZBs (blue), the BZQs (red) and the BZUs (green)
                        that belong to the LGB sample (see Section~\ref{sec:flux} for more details). 
                        {\it Right panel)} The trend between the radio and the gamma-ray luminosities for the subsample of LGB blazars
                        with a firm redshift estimate as reported in the \bzcat\ \citep{massaro10a,massaro11}. { The regression line is also shown in black.}}
          \label{fig:fluxes}
          \end{figure*}

In Figure~\ref{fig:fluxes}, we report the scatter plot of the $\gamma$-ray $vs$ the radio flux densities (i.e., $S_{\gamma}$ $vs$ $S_{325}$),
similar to the one in the 2LAC catalog \citep{ackermann11a}.
The $\gamma$-ray flux density, found in the 2LAC catalog \citep{ackermann11a}, is in units of photons cm$^{-2}$ s$^{-1}$ MeV$^{-1}$
while the radio flux densities were extracted from the WENSS catalog \citep{rengelink97}. 
As occurred for the 2LAC blazars and for the $\gamma$-ray blazars associated 
with a \wse\ counterpart, there is a good match between the WENSS and the \fer\ survey,
since sources that are bright in the radio (i.e., at 325 MHz) are generally bright also in the $\gamma$-rays \citep[e.g.,][]{ackermann11b}.
For the sources in the LGB for which a redshift estimate (i.e., 81 BZQs, 33 BZBs and 12 BZUs) 
was firmly established we also present the radio $vs$ $\gamma$-ray luminosity plot.
The radio luminosity have been estimated as 4$\pi\,D_L^2\,\nu\,S_{\nu}$ evaluated at 325 MHz,
where $D_L$ is the luminosity distance.
There is a clear correlation between { the luminosities, having the Spearman's rank and
and the Pearson correlation coefficients values of 0.85 and 0.91,  respectively. However,} selection effects 
could drive this trend since the subsample of the LGB constituted by blazars with redshift estimates
is neither complete or flux limited. On the other hand, Figure~\ref{fig:fluxes} 
again supports the flux match between the WENSS and the \fer\ surveys.

{ We remark that only 874 out 1143 \bzcat\ sources have a unique correspondence in the WENSS survey
within a radius of 8\arcsec.5 (total associations), thus 269 sources ($\sim$24\%) of them were nor associated or either detected at 325 MHz.
Given the distributions of the radio spectral indices for the LB sample, we extrapolated their 1.4 GHz flux densities down to 325 MHz,
to verify if these extrapolations are consistent with the completeness limit of the WENSS survey of $\sim$30 mJy \citep{rengelink97}.
We assumed values of $\alpha_{325}^{1400}$ equal to 0.25 for the BZBs, 0.08 for the BZQs and 0.21 for the BZUs,
as of the peak of their distributions (see Figure~\ref{fig:hist_bzcat}) and
we computed the $S_{325}$ flux densities from the NVSS $S_{1400}$ values reported in the \bzcat.

We found that 121 sources out of 269 without WENSS counterpart within 8\arcsec.5'' have the extrapolated flux 
densities below the threshold of 30 mJy, thus implying that the fraction of sources 
not observed in WENSS, because too faint to be detected, is small.
Moreover, we note that one of the reason underlying this lack of associations, could be due to a conservative choice of
our association radius $R_A$ (see Section~\ref{sec:radius}).
Thus increasing $R_A$ up to 20\arcsec\ we found additional 83 WENSS correspondences: 47 for the BZBs, 
28 for the BZQs and 8 for the BZUs, in particular, 56 of them among those expected to have extrapolated flux above the survey limit,
decreasing the number of blazars with an expected WENSS counterpart to 65 sources.
 
We also computed the $\alpha_{325}^{1400}$ for these new associations found within 8\arcsec.5 -20\arcsec\ 
$R_A$ range and we found that all their spectral indices 
are in agreement with the previous distributions of the LB sample, thus suggesting that these are reliable associations.
For our analysis, we preferred to use a more conservative approach keeping 
our association radius $R_A$ equal to 8\arcsec.5, so not modifying the samples previously defined, 
since the modest fraction of additional associations found does not affect our investigation.

Several effects or their combination might lead to a non-detection of the remaining 65 blazars within a radius of 20\arcsec\
as: fitting problem of the WSRT pipeline, confusion with nearby bright radio sources, intrinsic source variability, 
free-free or synchrotron self absorption. Then an additional possible explanation could be a too optimistic 
choice of the spectral index that we used to calculate the extrapolated flux values
Moreover, we could also expect that the spectral index of the remaining undetected 65 of blazars in the WENSS footprint,
is, on average, flatter that the average LB distribution otherwise it would have been easier to detect them,
since a the WENSS survey tends to observe steep sources.

Thus the limited fraction (i.e. $\sim$24\%) of the blazars lying in the footprint of the WENSS survey 
was not associated within 8\arcsec.5, decreasing to $\sim$18\% when increasing $R_A$ to 20\arcsec,
implies that the average behavior of the whole population is well represented by the current LB sample.}

\section{Unidentified Gamma-ray Sources}
\label{sec:ugs}

\subsection{Association procedure}
\label{sec:assoc}
Since $\gamma$-ray blazars appear to be associated with WENSS sources, 
we explore the possibility to associate radio sources detected in both the WENSS and the NVSS
with the UGSs.
The complete sample of UGSs that do not present any $\gamma$-ray analysis flag lists 299 sources,
however only 65 of them lie at declination larger than $\sim$+28\degr\ and could be considered for our investigation.

First we searched for all the WENSS sources that lie within a circular region of radius $\theta_{95}$,
corresponding to the semi-major axis of positional uncertainty ellipse at 95\% level of confidence,
centered on the $\gamma$-ray position of the UGSs listed in the 2FGL \citep{nolan12},
so defining our {\it search region} \citep[see also][]{paper4,paper6}.
Second, for each WENSS radio source we searched for the counterpart 
in the NVSS archive within a radius of 8\arcsec.5 and we computed 
the radio spectral indices $\alpha_{325}^{1400}$, according to Eq. (1).
{ Then, we indicated as potential blazar-like counterpart of each UGS analyzed 
the positionally closest radio source with a flat spectral index in agreement 
with that of the LGB sample. We also considered additional multifrequency information,
whenever present, that could confirm the blazar-like nature of our candidates
(see Appendix for more details).}

We distinguish between $\gamma$-ray blazar candidate of type A, 
radio sources having -1.00$\leq\alpha_{325}^{1400}\leq$0.55
and type B those with 0.55$<\alpha_{325}^{1400}<$0.65, since
as occurs for 90\% of the blazars in the LGB sample $\alpha_{325}^{1400}\leq$0.65.
{ We highlight that type A candidates are more reliable to be blazar-like sources
with respect to type B ones, since they show spectral indices in agreement with the largest fraction 
of LGB sources.}

This association procedure is based on the physical property that blazars show radio flat spectral indices 
also when it is calculated considering the low radio frequency at 325 MHz as occurs when using high frequency radio observations
\citep[e.g., Combined Radio All-Sky Targeted Eight-GHz Survey catalog, CRATES;][]{healey07}. 
The simplest physical interpretation of this fact could reside in the blazar nature, since they are core dominated radio sources,
even at low radio frequency their flux could be strongly dominated by the inner jet rather than by { large-scale} structures as for example
occurs in lobes of radio galaxies \citep[e.g.][]{blandford78b,blandford79}.
Moreover, it is worth noting that the large fraction of the sources in the LB sample are single-component sources
and this fraction is smaller in the LGB sample where only 12 objects out of 225 show multi-components at low radio frequencies.

We found that only 32 out of 65 UGSs have at least one WENSS that lie in the {\it search region}
with a NVSS or a FIRST counterpart within 8\arcsec.5 distance from the low frequency radio position;
for 18 WENSS this correspondence is unique while additional 14 UGSs have multiple matches, for a total of 58 WENSS radio sources
that lie within each {\it search region} of radius $\theta_{95}$.
We only considered the WENSS sources that also have a counterpart in the NVSS or in the FIRST catalogs
because the $S_{1400}$ flux density is necessary to derive the $\alpha_{325}^{1400}$ and apply our association method.
The radio spectral indices $\alpha_{325}^{1400}$ computed for the selected candidates are reported in Table~\ref{tab:main}.

\subsection{Correlation with existing databases}
\label{sec:counterparts}
We searched several major IR and optical surveys
for any possible counterpart within 3.3 \arcsec\ from the NVSS or the FIRST 
positions of our $\gamma$-ray blazar candidates, to verify if additional information can confirm their blazar-like nature.
The angular separation of 3.3 \arcsec\ has been chosen on the basis of 
the statistical analysis performed to associate each source of \bzcat\ with its IR counterpart
in the all-sky survey recently performed by \wse\ \citep{wright10}, 
so corresponding to the best searching radius from the 
NVSS position that is the one reported in the \bzcat\ \citep[see][for more details]{paper6}.
          \begin{figure*}[] 
           \includegraphics[height=9.5cm,width=6.5cm,angle=-90]{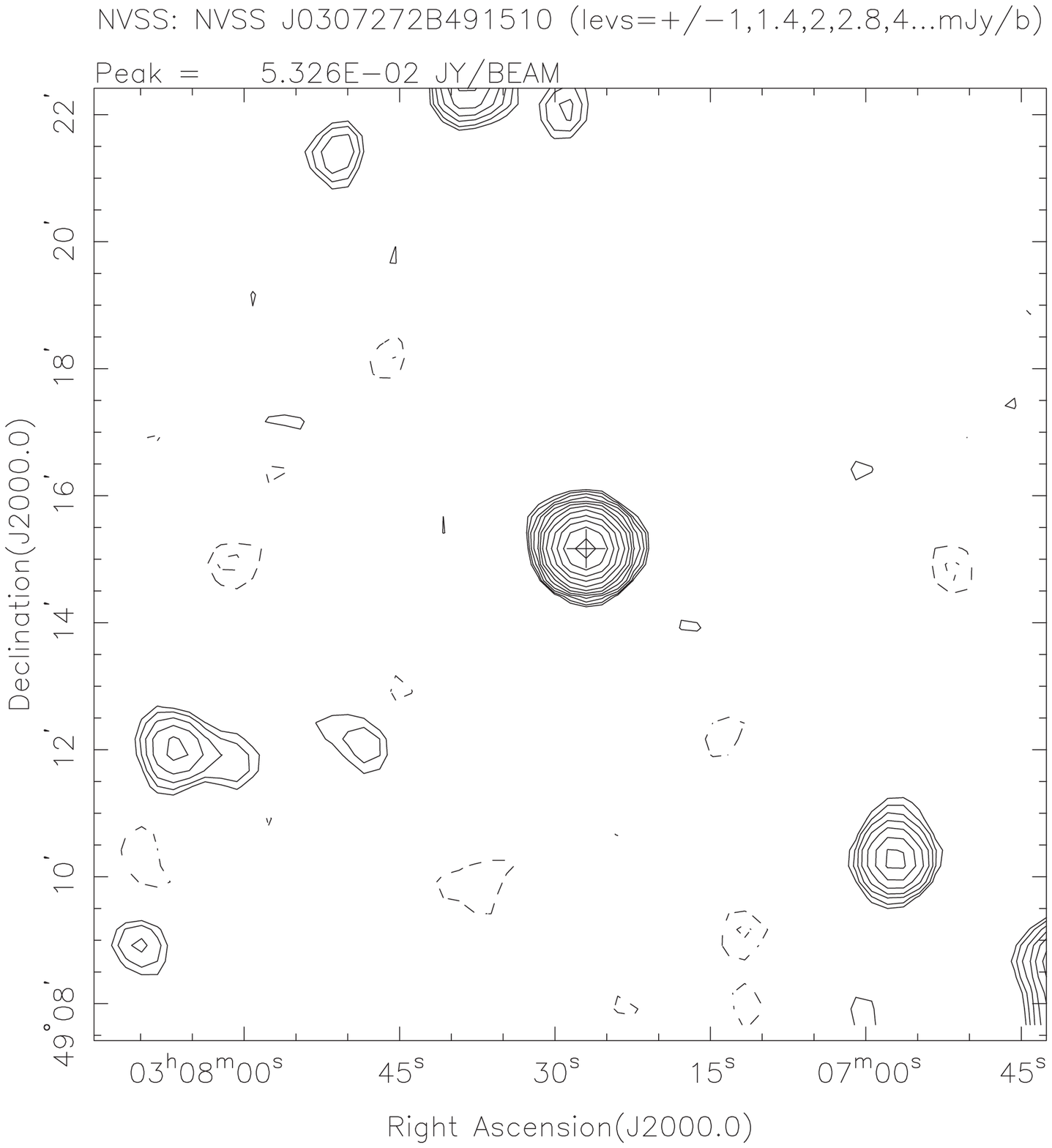}
           \includegraphics[height=9.5cm,width=6.5cm,angle=-90]{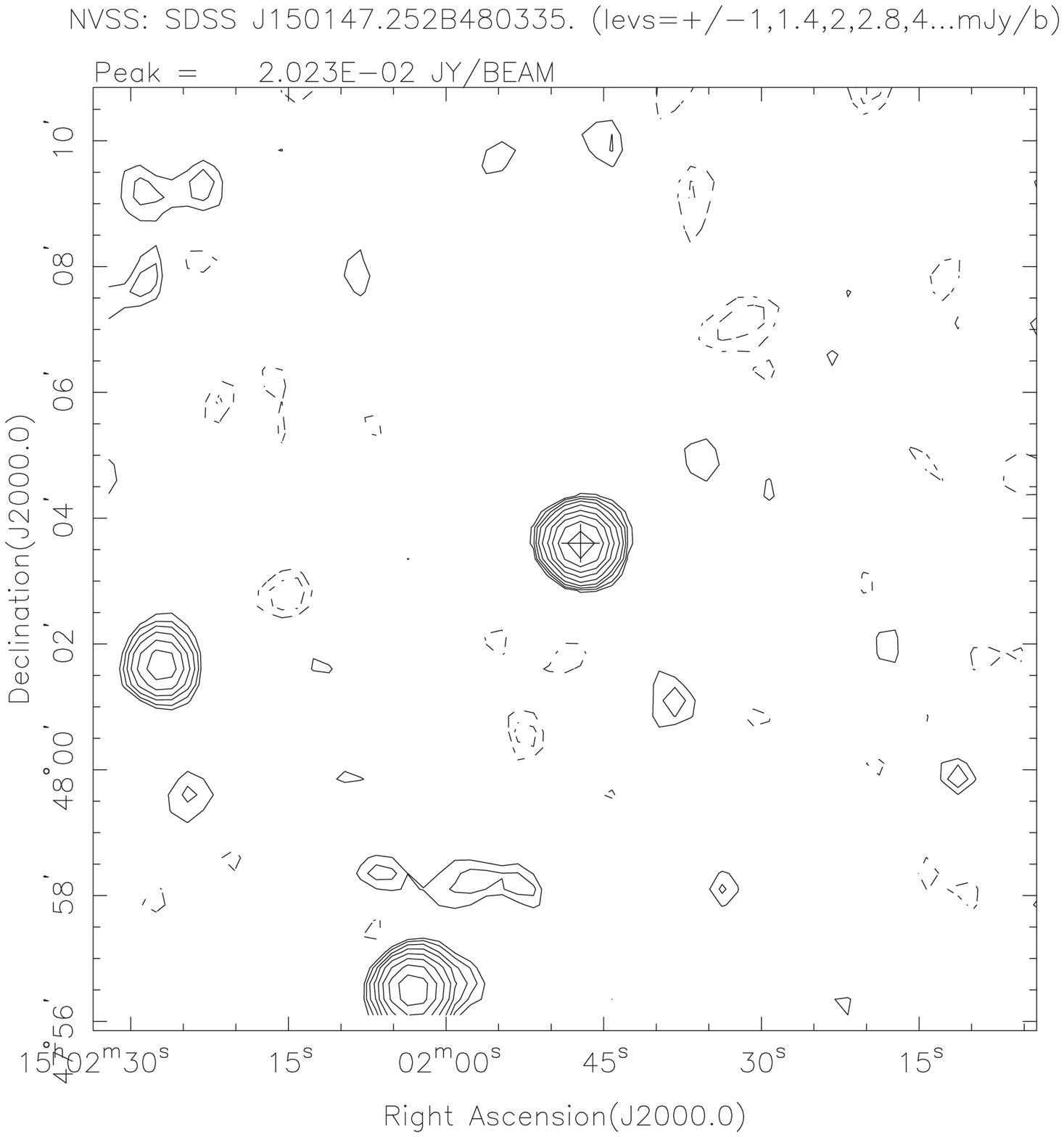}
          \caption{The archival NVSS radio observations (15\arcmin\ radius)
                        of the $\gamma$-ray blazars candidates: WN 0303.9+4903 (left panel) and WN 1500.0+4815 (right panel) uniquely
                        associated with the \fer\ sources 2FGL J0307.4+4915 and 2FGL J1502.4+4804, respectively. 
                        The black crosses point to the radio counterpart of the $\gamma$-ray blazar candidates selected 
                        according to our association procedure.
                        They are clear examples of core dominated radio sources similar to blazars at 1.4 GHz.
                        Contour levels are also reported together with the NVSS peak flux in units of Jy/beam.}
          \label{fig:radio}
          \end{figure*}

For the IR catalogs, we used the archival observations of the \wse\
that mapped the whole sky at 3.4, 4.6, 12, and 22 $\mu$m \citep{wright10} 
together with the Two Micron All Sky Survey \citep[2MASS - M;][]{skrutskie06}
since each \wse\ source is already associated with the closest 2MASS object 
{ by default in the \wse\ catalog} \citep[see][for more details]{cutri12}.
In addition we searched for optical counterparts, with possible spectra available, 
in the Sloan Digital Sky Survey \citep[SDSS dr9 - s; e.g.][]{adelman08,paris12}, in the Six-degree-Field Galaxy Redshift Survey 
\citep[6dFGS - 6;][]{jones04,jones09}. Then, we also considered 
the NASA Extragalactic Database (NED)\footnote{\underline{http://ned.ipac.caltech.edu/}} for other multifrequency information.
Finally, we cross correlate our sample with the USNO-B Catalog \citep{monet03} to identify the optical counterparts of our 
$\gamma$-ray blazar candidates; 
this is important to prepare and plan future follow up observations (see Table~\ref{tab:usno}). 
We found only 16 unique correspondences out of 22
between USNO-B and the radio positions of the selected $\gamma$-ray blazar candidates.

In Table~\ref{tab:main} we summarize the results of our multifrequency investigation,
reporting the 2FGL source name, together with that of the WENSS associated counterpart
and the NVSS name, the $\alpha_{325}^{1400}$ derived using Eq. (1), IR colors of the \wse\ counterpart, 
whenever present within 3.3\arcsec\ from the NVSS position, and 
together with several notes regarding the multifrequency archival analysis
{ (e.g., optical classification, redshift, etc.)}.

In case of multiple WENSS sources lying within the positional uncertainty of a single UGS
at 95\% level of confidence, the source having the flattest spectral index was considered the most probable counterpart.
We found 22 $\gamma$-ray blazar candidates, 14 of type A and 8 of type B, 
out of 32 UGSs selected with at least a WENSS radio source as candidate counterpart.
In addition, we also consider as $\gamma$-ray blazar candidate the WENSS source WN 1514.8+3701
that even if presenting a steep radio spectrum $\alpha_{325}^{1400}$ = 0.73 is variable in its radio emission 
\citep[see][for the 1.4 GHz flux measurements]{becker95,condon98},
for a total of 23 potential counterparts out of the complete sample of 65 UGSs investigated.

Source details derived from the multifrequency analysis are discussed in the following section, while
the complete list of USNO-B correspondences with their optical magnitudes is reported in Appendix.\\ 
In Figure~\ref{fig:radio} we also show the 15\arcmin\ radius NVSS images of two $\gamma$-ray blazar candidates, 
selected with the combination of WENSS and the 1.4 GHz surveys (i.e., NVSS and FIRST) 
to show that they are clearly compact and core dominated radio sources as expected for $\gamma$-ray blazar candidates 
\citep[e.g.][for a recent review]{massaro11}.

\begin{figure}[] 
           \includegraphics[height=9.5cm,width=8cm,angle=-90]{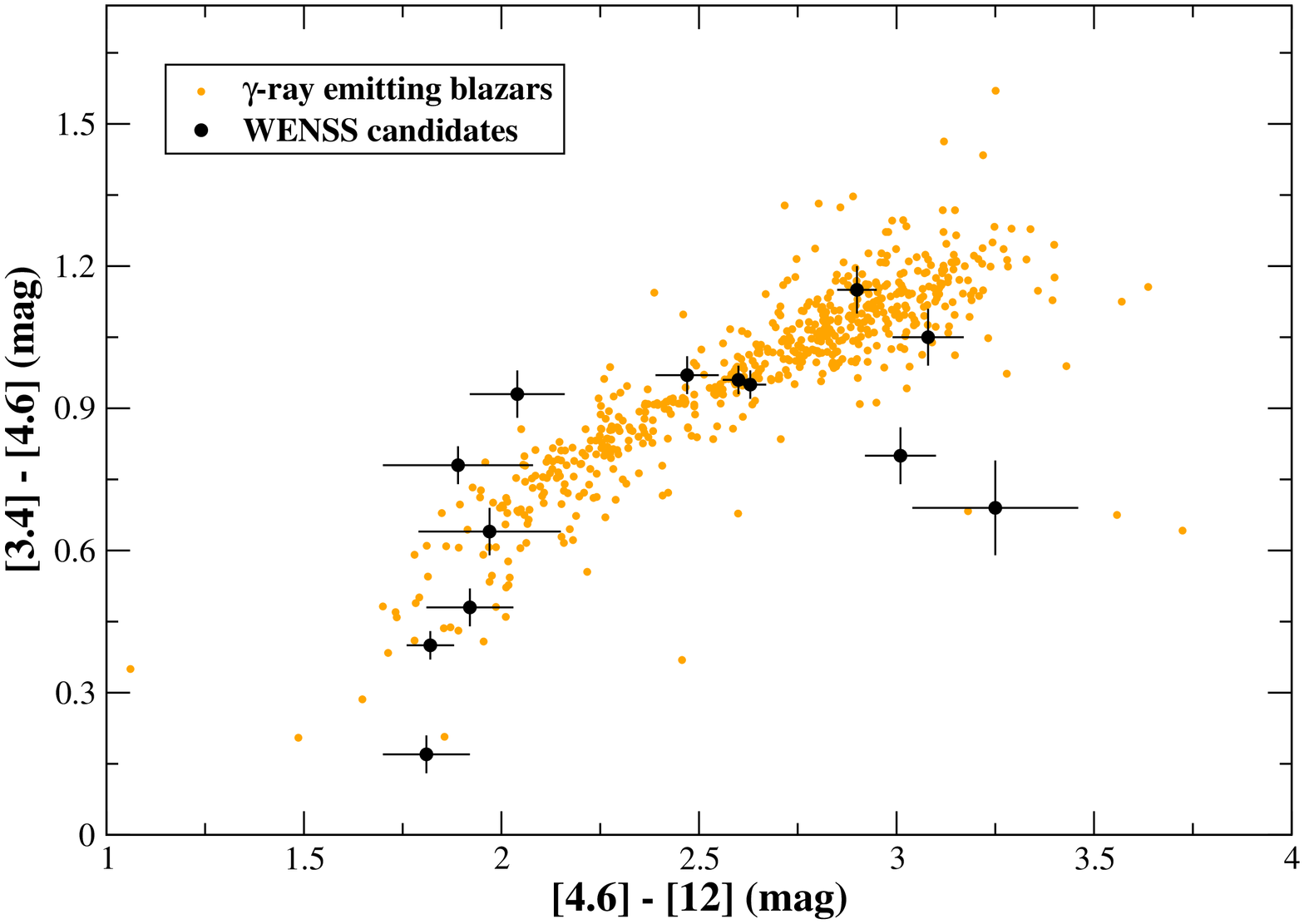}
          \caption{The [3.4]-[4.6]-[12] $\mu$m color-color plot of all the \wse\ counterparts of the WENSS-$\gamma$-ray blazar candidates (black circles),
                        detected at least in the first three \wse\ bands at 3.4, 4.6 and 12 $\mu$m in comparison with the blazars that constitute the 
                        \wse\ Gamma-ray strip \citep{paper2,paper3,paper6}. WENSS sources with IR counterparts only detected at 3.4 and 4.6 $\mu$m are not reported.
                        It is clear that the two samples have similar IR properties.}
          \label{fig:plane1}
          \end{figure}

Finally, in Figure \ref{fig:plane1} we show the comparison between the IR \wse\ colors of 
$\gamma$-ray emitting blazars and those of the candidates selected in this work based on the WENSS and the NVSS radio observations.
We only report the [3.4]-[4.6]-[12] $\mu$m color-color projection because not all the sources are detected at 22$\mu$m \citep[see][for more details]{paper6}.
IR counterpart of the WENSS $\gamma$-ray blazar candidates that are only detected at 3.4 and 4.6 $\mu$m are not shown.
{ The large fraction of the WENSS $\gamma$-ray blazar candidates show IR colros consistent with those of the \wse\ Gamma-ray strip \citep{paper2,paper3,paper6},
as expected for the $\gamma$-ray blazar candidates, strengthening our results.
However, we note that there are three WENSS sources lie in regions of the IR color-color plot less populated by $\gamma$-ray blazars
close to the boundaries of the \wse\ Gamma-ray strip.
In addition, these sources do not show any peculiarity in their radio emission above 325 MHz.}

\subsection{Spurious associations}
\label{sec:chance}
To estimate the probability that our associations are spurious
we considered the following approach.

First we created a {\it fake} $\gamma$-ray catalog of 216 positions,
shifting the coordinates of the 222 $\gamma$-ray blazars in the LGB sample by 0\degr.7 in a random direction of the sky 
within the WENSS footprint, keeping the same values of $\theta_{95}$, and verifying that no sources were detected by \fer\ 
within a circular region of radius $\theta_{95}$.
For each WENSS source within the positional uncertainty region at 95\% of confidence of the {\it fake} $\gamma$-ray catalog
we searched for the NVSS counterpart within a radius of 8\arcsec.5 and we calculated the radio spectral index $\alpha^{1400}_{325}$.
 
{ Then we establish the probability of spurious associations for type A and type B sources
concerning our procedure, by deriving the number of WENSS sources of type A and type B 
associated with the {\it fake} $\gamma$-ray catalog.
We found that there are 22 radio sources of type A and 15 of type B out of the 
216 fake objects, thus corresponding to 10\% and 7\% probabilities of spurious 
associations for type A and type B sources, respectively.

We emphasize that these estimates depend by the $\gamma$-ray background model,
the detection threshold and the flux limit of the 2FGL catalog \citep{nolan12},
in which no $\gamma$-ray emission is arising from 
the 22 sources of type A and the 15 of type B at the positions listed 
in the {\it fake} $\gamma$-ray catalog;

We note that, the probability of spurious associations 
for type A radio sources is larger than that for type B,
this is due to their range of $\alpha^{1400}_{325}$ chosen, 
larger in the former than in the latter type.
However, the type B sources could be more contaminated by 
non blazar-like radio sources as discussed in Section~\ref{sec:summary}.

As additional check on the above estimates on the probability 
of spurious associations, we considered the WENSS source density.}
The number of sources in the WENSS catalog above the limit of $\sim$30 mJy,
where the survey is almost complete \citep{rengelink97} is 135817 over an area of 3.1 sr (i.e. $\sim$10177 deg$^2$),
implying a source density $n_{wenss}$ of $\sim$13 sources per square degree.
Then, we searched for the NVSS counterparts of all the WENSS radio sources with $S_{325}>$30 mJy
within a radius of 8\arcsec.5, correspondent to our choice of $R_A$, and we found 109538 NVSS-WENSS associations.
For all of them, we computed the $\alpha^{1400}_{325}$ to estimate the source density of type A and type B sources.

We found that there are 17260 and 13491 radio sources of type A and type B in the footprint of the WENSS, respectively
corresponding to the following source density: $n_A \sim$ 2 src/deg$^2$ and $n_B \sim$ 1 src/deg$^2$, if assuming that both 
the NVSS and the WENSS surveys have uniform source densities and their combination over a scale radius $R_A$ 
has still a uniform source distribution.
Then, considering that the counterparts of 95\% of the $\gamma$-ray blazars listed in the 2LAC lie within an area of 0.01deg$^2$
at 95\% level of confidence, the probability to find a random WENSS radio source with a NVSS counterpart within 8\arcsec.5
within the \fer\ positional uncertainty is $\sim$2\% for sources of type A and $\sim$1\% for sources of type B, under the above assumptions.
These second estimates can be considered as lower limit of the expected number of spurious associations
for our $\gamma$-ray blazar catalogs.

\section{Comparison with other methods}
\label{sec:comparison}
We note that among the whole sample of 65 UGSs analyzed, 
there are 8 sources for which we found at least one $\gamma$-ray blazar candidate
that were also unidentified in the First Fermi $\gamma$-ray LAT catalog (1FGL) \citep{abdo10} and 
were analyzed using two different statistical approaches: the Classification Tree
and the Logistic regression analyses \citep[see][and references therein]{ackermann12}.
For these 8 UGSs, analyzed on the basis of the above statistical approaches, 
we performed a comparison with our results
to verify if the 2FGL sources that we associated with a $\gamma$-ray blazar 
candidates have been also classified as AGNs.

By comparing the results of our association method with those in Ackermann et al. (2012), we found that
7 out of 8 UGSs that we associate to a $\gamma$-ray blazar candidate are also classified as AGNs,
all of them with a probability higher than 60\% with 6 higher than 78\%. 
The remaining source was classified as pulsar candidates but with a very low probability (i.e. 60\%)
Consequently, we emphasize that our results are in good agreement with the classification 
suggested previously by Ackermann et al. (2012)
consistent with the $\gamma$-ray blazar nature of the \wse\ candidates proposed in our analysis.

We also compare our results with those of Massaro et al. (2013), 
that selected $\gamma$-ray blazar candidates for all the UGSs on the basis of the \wse\ IR colors \citep[see also][]{paper6}.
{ Among the sample of UGS analyzed in the present paper only 9 cases are also in the list of Massaro et al. (2013).
In particular, for 4 out of 9 UGSs having a $\gamma$-ray blazar candidate selected 
using the low-frequency radio spectral index, the assigned counterpart is the same as proposed by Massaro et al. (2013), 
increasing the reliability of our radio selections.
On the other hand, the remaining 5 WENSS candidates, potential counterpart of UGSs,
are not bright enough at 12$\mu$m and at 22$\mu$m to be detected by \wse.
Since the \wse\ association method based on the IR colors of $\gamma$-ray blazars
requires that all candidates must be detected in the four \wse\ bands \citep[see][for more details]{paper4,paper7}
it would automatically fail for these remaining 5 WENSS candidates. 
The possibility of missing some $\gamma$-ray blazar candidates 
using the \wse\ association procedure is highlighted by its completeness that is $\sim$97\% \citep{paper6}, suggesting
that additional methods, as the one developed here on low radio frequency observations, should to be used to 
find additional $\gamma$-ray blazar-like sources.}

\section{Discussion and conclusions}
\label{sec:summary}
We combined the archival observations of the WENSS \citep{rengelink97}
with those of the NVSS \citep{condon98} and of the FIRST catalogs \citep{becker95,white92}
to characterize the radio emission of the known blazars listed in the \bzcat\ \citep{massaro09,massaro11}.
Then, we investigated the distribution of the radio spectral index $\alpha_{325}^{1400}$ 
for the blazars with a counterpart in the WENSS focusing on those that are associated with $\gamma$-ray sources.
We found that as occurs using the high frequency radio data \citep[e.g.,][]{healey07}, 
about 80\% of $\gamma$-ray emitting blazars have flat radio spectra (i.e., $\alpha_{325}^{1400}<$0.5),
with 99\% even smaller than 0.75 (see Section~\ref{sec:index} for more details).
This strongly implies that at low radio frequencies blazar emission is still dominated by the beamed radiation arising from their jets,
and not from emission of extended structures as lobes, in agreement also with the recent results of Kimball et al. (2011).

On the basis of the distributions of the radio spectral index $\alpha_{325}^{1400}$ derived from the blazars in the footprint of both
the NVSS or the FIRST and the WENSS radio surveys we developed a new association procedure for the UGSs.
We searched for all the WENSS sources that lie within a {\it search region} defined as a circle of radius $\theta_{95}$,
corresponding to the semi-major axis of positional uncertainty ellipse at 95\% level of confidence
and centered on the $\gamma$-ray position of the UGSs listed in the 2FGL \citep{nolan12}.
We associated each WENSS source to its NVSS counterpart, whenever present
within 8\arcsec.5 angular separation, and we calculated the $\alpha_{325}^{1400}$.
Then, we considered as a candidate counterparts of each UGS, those radio sources with the radio
spectral index in agreement with those of the $\gamma$-ray blazars and with additional multifrequency information, when present,
confirming its blazar-like origin (see Section~\ref{sec:ugs} for more details). 

We distinguished between $\gamma$-ray blazar candidate of type A, radio sources having -1.00$\leq\alpha_{325}^{1400}\leq$0.55
and type B those with 0.55$<\alpha_{325}^{1400}<$0.65, since
as occurs for 90\% of the blazars in the LGB sample $\alpha_{325}^{1400}\leq$0.65 (see Section~\ref{sec:ugs} for more details).
A similar attempt has been recently used to search for flat spectrum radio sources 
as potential counterparts of the unidentified INTEGRAL sources \citep{maiorano11,molina12}.
However, their approach was based on different radio surveys and the low frequency radio
observations were indeed not used to estimate the radio spectral index, as proposed in our new procedure.

We found 22 $\gamma$-ray blazar candidates, 14 of type A and 8 of type B, out of 65 UGSs analyzed.
In addition, we suggested the WENSS source WN 1514.8+3701 also as potential counterpart of 2FGL J1517.2+364;
its steep radio spectrum does not fit in any of the previous type of candidates, however,  
it shows radio variability at 1.4 GHz, thus bringing the total number of $\gamma$-ray blazar candidates to 23 out of 65 UGSs.

Our new approach presented here has several advantages with respect to those previously 
adopted to associate $\gamma$-ray sources \citep[e.g.][]{hartman99,pittori09,nolan12}.
At low radio frequencies there is only a handful of source classes that are emitting, thus being potential contaminants of our selection
for $\gamma$-ray blazar candidates; these are: galaxy clusters, supernova remnants, pulsars (PSRs), 
radio galaxies { as} ultra steep spectrum radio sources (USSs) \citep[e.g.][]{rottgering94,miley08} { or}
compact steep spectrum radio sources (CSSs) and { Gigahertz-peaked spectrum (GPS) radio sources} 
\citep[e.g.,][]{saikia95,odea98,giroletti09a} and 
Seyfert galaxies \citep[e.g., ][for a recent review]{singh11}.

However, the first two source classes mentioned above are extended and combining the low frequency data with those at 
higher frequency can be easily excluded, or in the nearby cases, simply using the MHz observations.
All the other classes of sources have a steep radio spectra and are not selected according to our criteria 
(see Section~\ref{sec:ugs}), in particular, young radio galaxies as the CSSs \citep[see also][]{fanti85,odea90,stanghellini97} 
or high redshift radio galaxies as the USSs \citep[see][for a recent review]{miley08}
have radio spectral indices systematically higher than 0.5 or even 1 by class definition, respectively.
{ While GPS sources can not be excluded in principle, as their convex spectrum could mimic a flat power law when only two frequency are considered, 
we do not find any evidence of spectral curvature in the few cases where we have a third available frequency (typically at 5 GHz) and the IR colors of 
our candidates appear to be blazar-like. We can further exclude} Seyfert galaxies, that could be detected at 1.4 GHz frequencies \citep[e.g.][]{giroletti09b} 
but with a steep radio spectrum, typically higher 0.7 \citep[e.g.,][and references therein]{singh11}.
On the other hand, flat spectrum radio sources are generally associated with quasars 
in agreement with our selection \citep[e.g.,][]{terasanta01,ivezic02,kimball08}

Steep radio spectra are also characteristic of the low frequency emission from pulsars, 
that could have values of $\alpha_{325}^{1400}$ $\sim$ 2 - 3 \citep[e.g.,][]{sieber73}.
We highlight that pulsars, the second largest known population of $\gamma$-ray sources, given their steep radio spectra
can be also identifiable on the basis of their spectral shape, using the low frequency radio observations. 
One example is the pulsar PSR J0218+4232 associate to the \fer\ source 2FGL J0218.1+4233 
\citep[][see also the Public List of LAT-Detected Gamma-Ray Pulsars 
\footnote{\underline{https://confluence.slac.stanford.edu/display/GLAMCOG/Public\\+List+of+LAT-Detected+Gamma-Ray+Pulsars}}]{nolan12},
and correspondent to the WENSS radio object WN 0214.9+4218, with a counterpart at 1.4 GHz  
and $\alpha_{325}^{1400}=$ 3 \citep[see][for additional details]{kouwenhoven00}.

\begin{figure*}[] 
           \includegraphics[height=9.5cm,width=6.5cm,angle=-90]{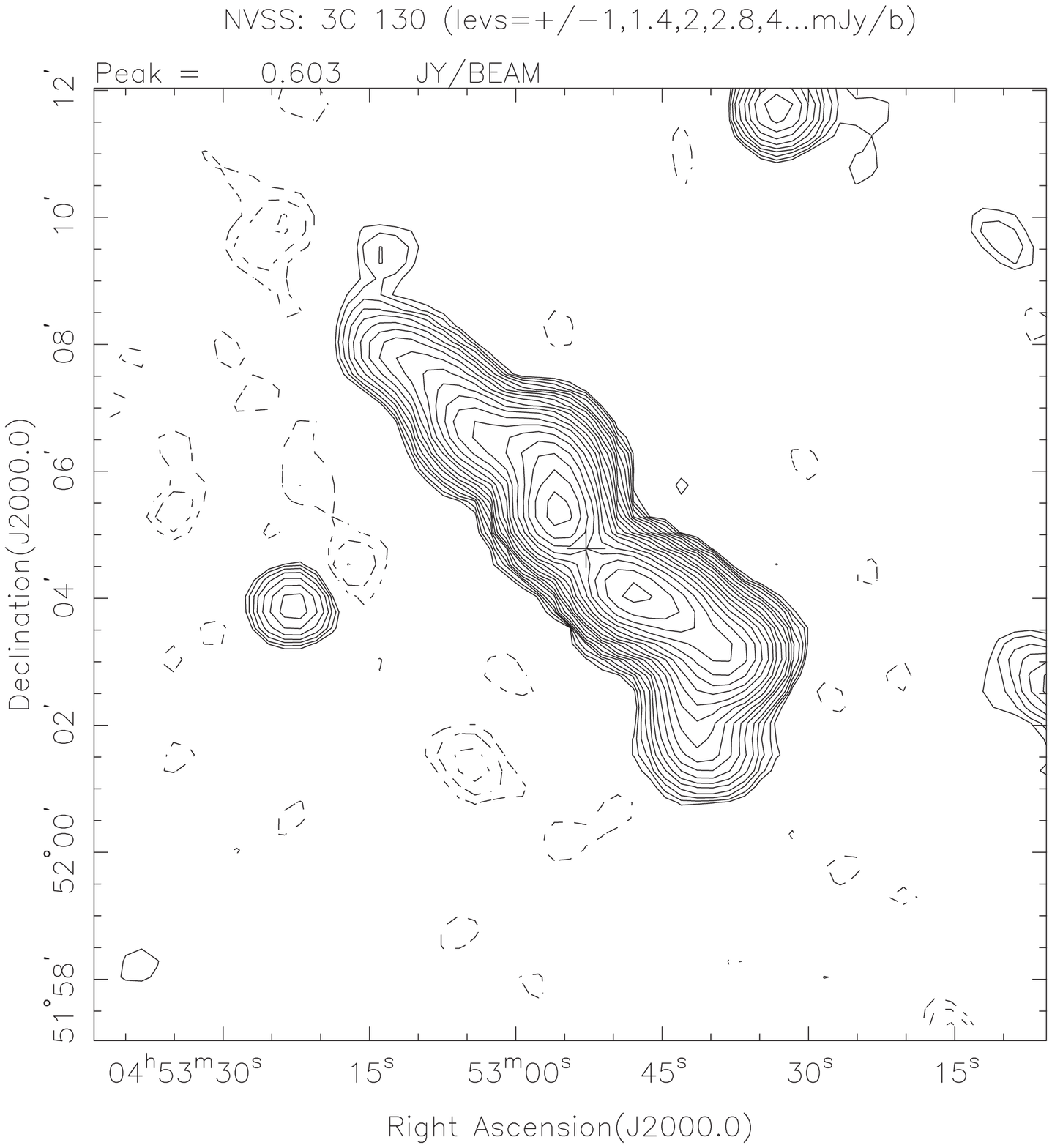}
           \includegraphics[height=9.5cm,width=6.5cm,angle=-90]{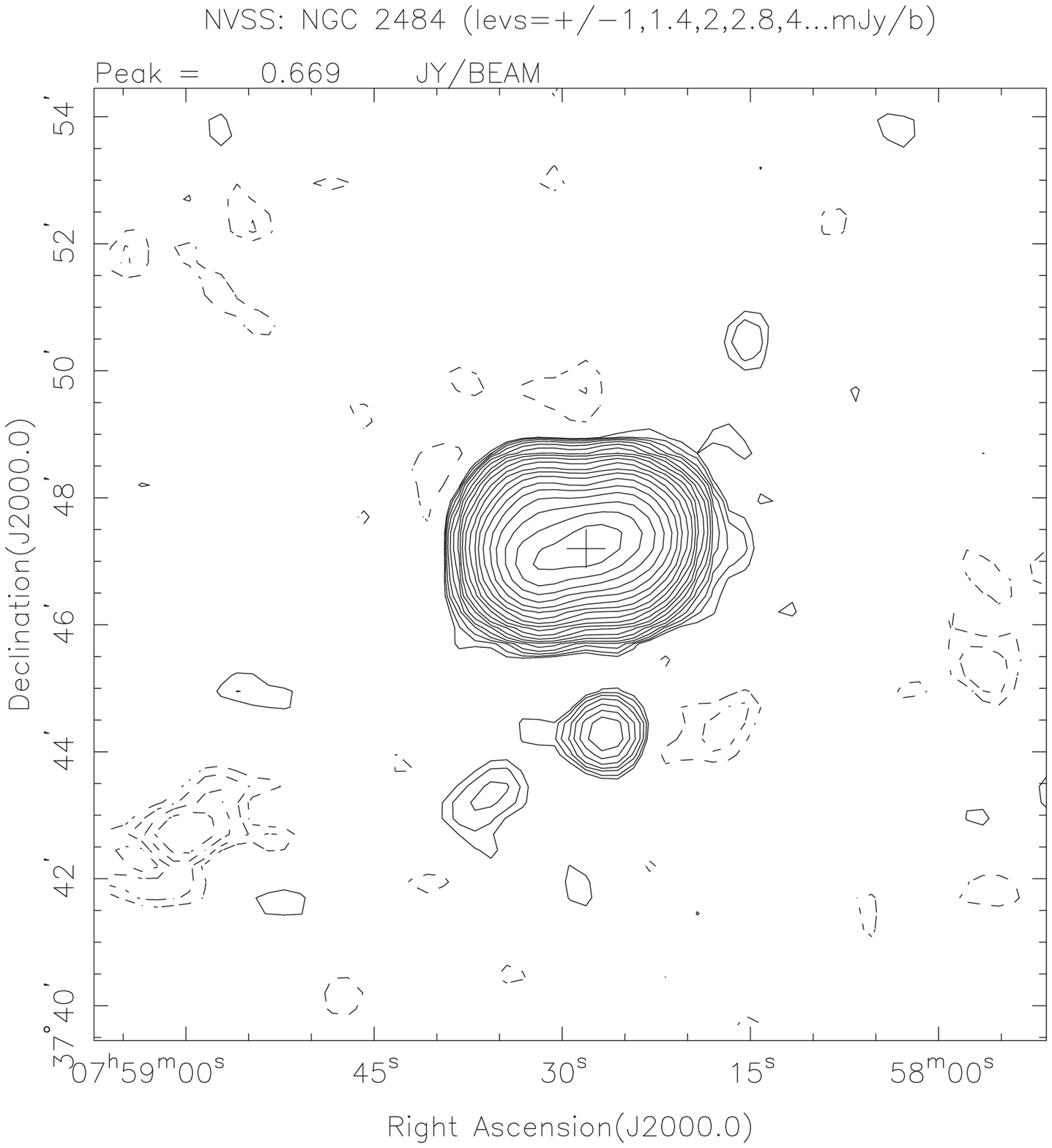}
           \includegraphics[height=9.5cm,width=6.5cm,angle=-90]{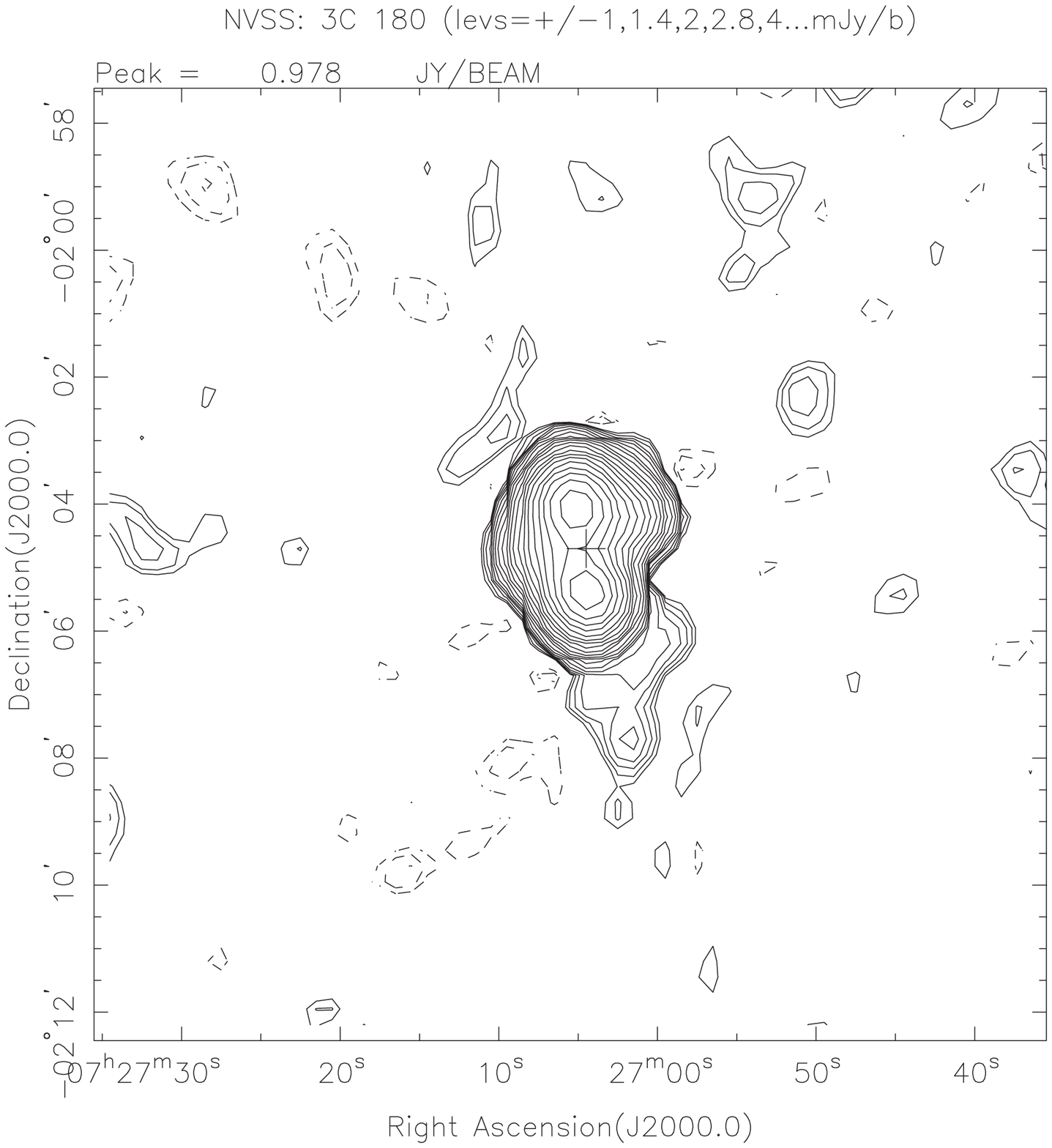}
           \includegraphics[height=9.5cm,width=6.5cm,angle=-90]{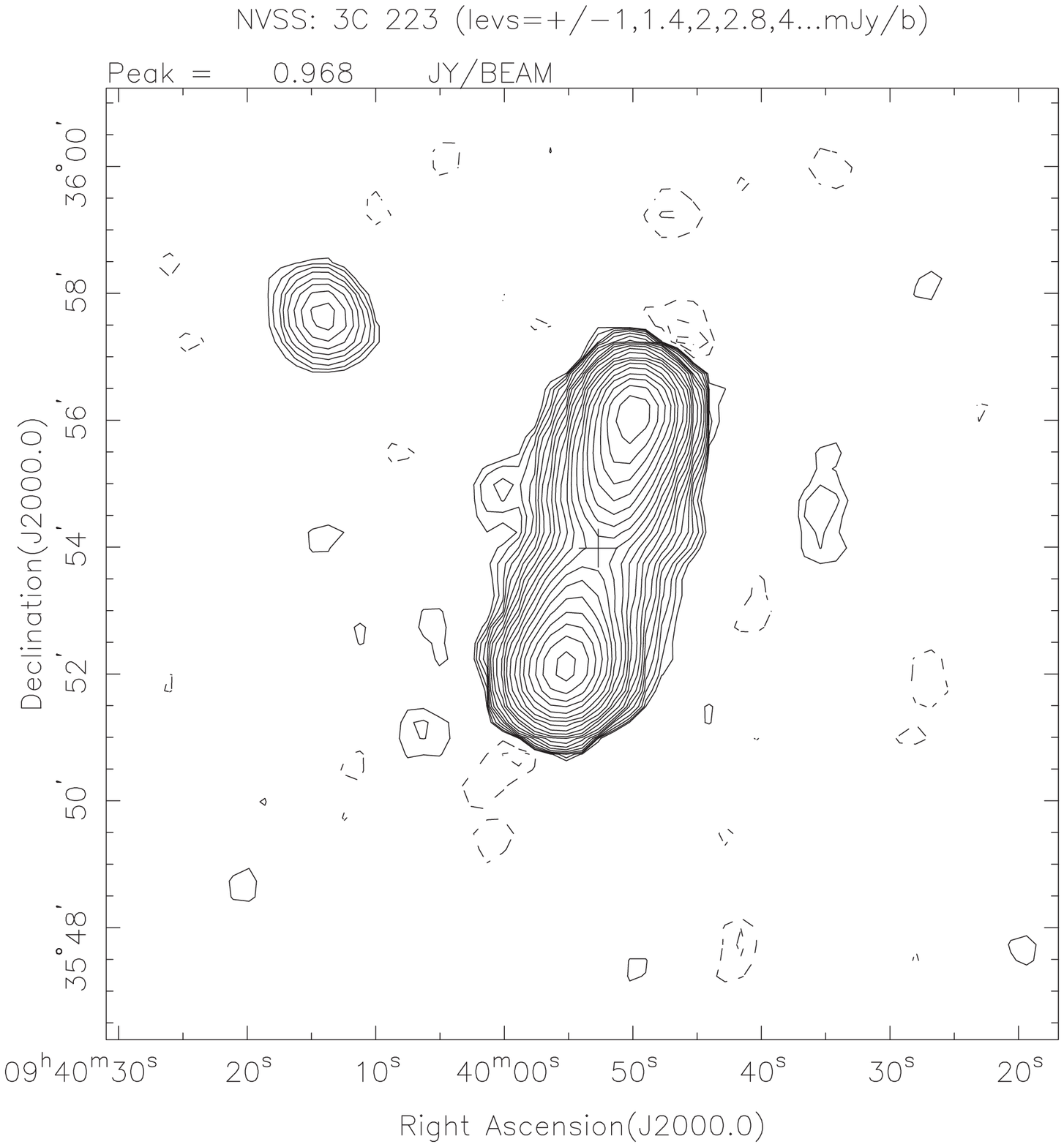}
           \caption{The archival NVSS radio images of four radio galaxies in the 3C catalog \citep{edge59}: 3C\,130 (upper left) and 3C\,189, alias NGC\,2484 (upper right)
                        as examples of FR\,Is while 3C\,180 (lower left) and 3C\,223 (lower right) as typical FR\,IIs.
                        It is clear how these sources could be distinguished by the core dominated ones selected according to our association procedure, 
                        not only considering their radio spectral index but also their radio morphology, at the least in the nearby cases
                        since their emission at 1.4 GHz is already dominated by that arising from extended structures.}
          \label{fig:rgs}
          \end{figure*}

{ Thus the only significant class of contaminants} of our association method could be the radio galaxies, 
parent population of blazars \citep[e.g.,][]{blandford78b,urry95}.
{ However, it is possible to disentangle between radio galaxies and blazars with the following two approaches
based on radio observations. 
Blazars being core dominated sources are unresolved by the NVSS and the WENSS observations,
while radio galaxies, in particular those lying at low redshifts, would appear extended, because
they are generally dominated by lobe emission even at 1.4 GHz, as for example shown in the four NVSS
images of typical FR\,I and FR\,IIs \citep{fanaroff74} radio sources (see Figure~\ref{fig:rgs}) that
belonging to the Third Cambridge Catalog of radio sources \citep[3C;][]{edge59} 
and have been recently observed by the 3C \chn\ snapshot survey \citep{massaro10b,massaro12c}.
On the other hand, if radio galaxies are not resolved due to large NVSS and WENSS radio beams,
their extended structure, characterized by steep spectra, will dominate the radio emission below 1.4GHz
allowing us to distinguish them from $\gamma$-ray blazar candidates simply using their radio spectral index.}

{ In particular,} extensive investigations of the low frequency radio emission in radio galaxies 
have been carried out on the 3C catalog of radio sources \citep[3C;][]{edge59} as for example the analyses performed 
by Kellerman et al. (1968) and Pauliny-Toth et al. (1968) \citep[see also][]{kellerman69a,kellerman69b}.
Thus, according to { these} studies all radio galaxies and quasars in the 3C sample have radio spectral indices 
systematically higher than 0.6, being a marginal contaminants for our selection of type B candidates.

We remark that since the low radio frequency catalogs were never previously used to associate $\gamma$-ray sources,
our new association procedure, here proposed, as that created on the basis of the \wse\ all-sky survey \citep{paper4,paper6,paper7}, 
is complementary to the other methods developed or adopted in the 1FGL and in the 2FGL \citep{nolan12}.
It is also worth noting that $\gamma$-ray blazar candidates listed here, were not selected with the different 2FGL methods 
because our WENSS flat spectrum radio sources are generally fainter than the flux thresholds of high frequency radio catalogs
as the CRATES \citep{healey07}, thus increasing the chance to find high redshift blazars as highlighted in the case of WN 1500.0+4815.

\begin{table*}
\tiny
\caption{Unidentified Gamma-ray sources.}
\begin{tabular}{|lllcccclc|}
\hline
  2FGL  &  WENSS  &  NVSS  &$\alpha_{325}^{1400}$& [3.4]-[4.6] & [4.6]-[12] & [12]-[22] & notes & z  \\
  name  &  name      &  name  &                                   &     mag     &    mag     &   mag     &           &     \\
\hline
\noalign{\smallskip}
J0002.7+6220  &  WN 0000.3+6202  &  NVSS J000253+621917  &  1.28  & -0.02(0.05) &  1.05(0.47) &  4.54(0.49) &  N,w,M         &  ?    \\
J0039.1+4331  &  WN 0036.7+4320  &  NVSS J003928+433651  &  0.73  &  0.01(0.10) &  $<$2.24    &  $<$4.05    &  N,w           &  ?    \\
J0158.6+8558  &WN 0140.0+8546$^A$&  NVSS J014929+860114  &  0.37  &  ---        &  ---        &  ---        &  N             &  ?    \\
              &WN 0144.4+8546$^A$&  NVSS J015409+860117  &  0.39  &  ---        &  ---        &  ---        &  N             &  ?    \\
              &  WN 0144.4+8550  &  NVSS J015417+860502  &  0.81  &  ---        &  ---        &  ---        &  N             &  ?    \\
J0248.5+5131  &WN 0244.6+5114$^B$&  NVSS J024805+512719  &  0.63  &  ---        &  ---        &  ---        &  N             &  ?    \\
              &WN 0245.0+5118$^A$&  NVSS J024834+513116  &  0.36  &  0.93(0.05) &  2.04(0.12) &  2.59(0.33  &  N,w,M         &  ?    \\%wise
J0307.4+4915  &WN 0303.9+4903$^A$&  NVSS J030727+491510  &  0.24  &  0.40(0.03) &  1.82(0.06) &  2.29(0.20) &  N,w,M         &  ?    \\%wise
J0332.1+6309  &WN 0327.5+6258$^A$&  NVSS J033153+630814  &  0.26  &  0.96(0.03) &  2.60(0.04) &  1.96(0.11) &  N,w,M         &  ?    \\%wise
J0336.0+7504  &  WN 0330.3+7500  &  NVSS J033613+751015  &  1.13  &  ---        &  ---        &  ---        &  N             &  ?    \\%VLSS
J0353.2+5653  &WN 0349.1+5645$^A$&  NVSS J035309+565431  &  0.34  &  0.78(0.04) &  1.89(0.19) &  $<$2.38    &  N,w,M,rv      &  ?    \\%wise
              &  WN 0349.5+5642  &  NVSS J035332+565141  &  1.03  & -0.04(0.07) &  2.11(0.34) &  $<$3.21    &  N,w,M         &  ?    \\%VLSS
              &  WN 0350.0+5639  &  NVSS J035404+564827  &  0.95  &  ---        &  ---        &  ---        &  N             &  ?    \\
J0600.9+3839  &WN 0557.5+3838$^B$&  NVSS J060102+383828  &  0.65  &  0.97(0.04) &  2.47(0.08) &  2.62(0.17) &  N,w           &  ?    \\%wise+VLSS
J0644.6+6034  &WN 0640.0+6041$^B$&  NVSS J064435+603849  &  0.64  &  0.64(0.05) &  1.97(0.18) &  $<$2.41    &  N,w           &  ?    \\%wise
J0723.9+2901  &WN 0720.7+2905$^A$&  NVSS J072354+285930  &  0.53  &  1.15(0.05) &  2.90(0.05) &  2.40(0.11) &  N,F,w         &  ?    \\%wise
J0843.6+6715  &WN 0839.4+6724$^A$&  NVSS J084403+671350  &  0.46  &  ---        &  ---        &  ---        &  N             &  ?    \\
              &  WN 0839.7+6728  &  NVSS J084420+671709  &  0.11  & -0.30(0.17) &  $<$2.94    &  $<$3.81    &  N,w           &  ?    \\
J0844.9+6214  &  WN 0839.8+6230  &  NVSS J084404+622010  &  0.98  &  ---        &  ---        &  ---        &  N,F           &  ?    \\
J0900.9+6736  &  WN 0856.1+6754  &  NVSS J090038+674223  &  0.68  &  0.58(0.14) &  4.22(0.17) &  $<$2.26    &  N,w           &  ?    \\%wise+VLSS
J1013.6+3434  &WN 1010.4+3445$^B$&  NVSS J101323+343039  &  0.59  &  0.08(0.24) &  $<$3.26    &  $<$3.53    &  N,F,w,s       &  ?    \\
J1016.1+5600  &WN 1012.4+5605$^B$&  NVSS J101544+555100  &  0.62  &  1.05(0.06) &  3.08(0.09) &  2.57(0.25) &  N,F,w,s       &  ?    \\%wise+VLSS
              &  WN 1012.5+5613  &  NVSS J101552+555844  &  0.86  &  ---        &  ---        &  ---        &  N             &  ?    \\
              &  WN 1013.0+5622  &  NVSS J101624+560702  &  0.66  &  -0.61(0.43)  & 4.73(0.50) &  $<$2.85    &  N,F,w,s         &  ?    \\
              &  WN 1013.3+5629  &  NVSS J101640+561445  &  1.15  &  ---        &  ---        &  ---        &  N,s           &  ?    \\%VLSS
              &WN 1013.6+5616$^B$&  NVSS J101657+560112  &  0.62  &  ---        &  ---        &  ---        &  N,F,s         &  ?    \\
              &  WN 1014.2+5616  &  NVSS J101734+560135  &  0.22  & -0.12(0.16) &  $<$3.25    &  $<$3.49    &  N,F,w,s       &  ?    \\
              &  WN 1014.3+5612  &  NVSS J101735+555738  &  0.95  &  0.50(0.06) &  3.30(0.10) &  2.40(0.28) &  N,F,w,M,s     &  ?    \\%wise
J1223.3+7954  &WN 1222.0+8010$^A$&  NVSS J122358+795329  &  0.22  &  0.48(0.04) &  1.92(0.11) &  2.23(0.38) &  N,w,M         &  ?    \\%wise
J1502.1+5548  &WN 1501.0+5603$^A$&  NVSS J150229+555204  &  0.25  &  ---        &  ---        &  ---        &  N,F           &  ?    \\
              &WN 1500.0+5556$^B$&  NVSS J150126+554442  &  0.64  &  ---        &  ---        &  ---        &  N             &  ?    \\
              &WN 1501.2+5559$^B$&  NVSS J150238+554810  &  0.60  &  0.46(0.11) &  2.73(0.35) &  3.21(0.61) &  N,F,w,s       &  ?    \\
J1502.4+4804  &WN 1500.0+4815$^A$&  NVSS J150147+480335  &  0.18  &  0.69(0.10) &  3.25(0.21) &  $<$3.23    &  N,F,w,s,QSO       &  2.78 \\%wise
              &  WN 1500.3+4808  &  NVSS J150203+475629  &  1.74  &  0.24(0.13) &  $<$3.12    &  $<$3.35    &  N,F,w         &  ?    \\
J1517.2+3645  &  WN 1514.6+3653  &  NVSS J151634+364217  &  1.01  &  ---        &  ---        &  ---        &  N,F,s         &  ?    \\
              &WN 1514.8+3701$^*$&  NVSS J151649+365023  &  0.73  &  0.95(0.03) &  2.63(0.04) &  2.07(0.12) &  N,F,w,s,rv,v  &  ?    \\%wise
              &  WN 1515.6+3701  &  NVSS J151732+365046  &  0.71  &  0.42(0.07) &  2.26(0.28) &  $<$3.49    &  N,F,w,s       &  ?    \\
              &  WN 1515.8+3657  &  NVSS J151744+364629  &  1.09  &  0.79(0.12) &  3.26(0.24) &  $<$3.28    &  N,w,s         &  ?    \\
J1614.8+4703  &WN 1612.6+4714$^B$&  NVSS J161408+470705  &  0.57  &  ---        &  ---        &  ---        &  N,F,s         &  ?    \\
              &  WN 1612.8+4712  &  NVSS J161422+470518  &  1.03  &  ---        &  ---        &  ---        &  N,s           &  ?    \\
              &  WN 1613.5+4711  &  NVSS J161505+470359  &  0.78  &  ---        &  ---        &  ---        &  N,F,s         &  ?    \\
J1623.2+4328  &  WN 1621.7+4341  &  NVSS J162323+433430  &  0.70  &  0.28(0.09) &  $<$2.69    &  $<$3.04    &  N,F,w,s       &  ?    \\
J1730.8+5427  &  WN 1730.4+5431  &  NVSS J173133+542924  &  2.08  &  ---        &  ---        &  ---        &  N,F,w,M,s     &  0.238    \\
              &  WN 1731.3+5435  &  NVSS J173221+543338  &  0.88  &  ---        &  ---        &  ---        &  N,s           &  ?    \\
J1738.9+8716  &  WN 1758.4+8718  &  NVSS J173722+871744  &  0.15  & -0.07(0.11) &  3.00(0.31) &  3.64(0.42) &  N,w,M         &  ?    \\
J1748.8+3418  &  WN 1747.3+3426  &  NVSS J174911+342528  &  0.73  &  0.25(0.04) &  $<$0.93    &  $<$3.61    &  N,w,M         &  ?    \\
              &WN 1747.6+3424$^B$&  NVSS J174929+342311  &  0.55  &  0.10(0.28) &  $<$3.40    &  $<$3.93    &  N,w           &  ?    \\
J1828.7+3231  &  WN 1826.1+3228  &  NVSS J182801+323005  &  0.84  &  ---        &  ---        &  ---        &  N             &  ?    \\
              &  WN 1826.7+3229  &  NVSS J182835+323108  &  0.84  &  ---        &  ---        &  ---        &  N             &  ?    \\
              &WN 1826.8+3220$^A$&  NVSS J182843+322248  &  0.53  &  0.28(0.13) &  $<$2.63    &  $<$4.36    &  N,w           &  ?    \\
J2041.2+4735  &  WN 2039.6+4728  &  NVSS J204119+473855  &  1.00  &  ---        &  ---        &  ---        &  N             &  ?    \\%VLSS
J2107.8+3652  &WN 2106.1+3643$^A$&  NVSS J210805+365526  &  -0.08 &  0.80(0.06) &  3.01(0.09) &  1.76(0.51) &  N,w           &  ?    \\%wise
              &  WN 2105.7+3645  &  NVSS J210747+365716  &   1.00 &  ---        &  ---        &  ---        &  N,M             &  ?    \\%wise
J2110.3+3822  &  WN 2108.3+3804  &  NVSS J211020+381659  &  0.68  &  0.17(0.04) &  1.81(0.11) &  2.15(0.40) &  N,w,M         &  ?    \\%wise
J2114.1+5440  &  WN 2113.6+5429  &  NVSS J211512+544221  &  0.70  &  ---        &  ---        &  ---        &  N             &  ?    \\
J2133.9+6645  &WN 2133.6+6625$^A$&  NVSS J213443+663847  &  0.54  &  ---        &  ---        &  ---        &  N             &  ?    \\
J2231.0+6512  &WN 2229.1+6451$^A$&  NVSS J223048+650658  &  0.05  &  ---        &  ---        &  ---        &  N             &  ?    \\
              &  WN 2230.2+6450  &  NVSS J223155+650556  &  1.01  &  ---        &  ---        &  ---        &  N             &  ?    \\%VLSS
\noalign{\smallskip}
\hline
\end{tabular}\\
Col. (1) 2FGL name. \\
Col. (2) WENSS name. The capital letter indicates $\gamma$-ray blazar candidate of type A or type B, respectively (see Section~\ref{sec:ugs} for details).\\
Col. (3) NVSS name. \\
Col. (4) radio spectral index $\alpha_{325}^{1400}$. \\
Cols. (5,6,7) IR colors from \wse. Values in parentheses are 1$\sigma$ uncertainties. \\
Col. (8) Notes: N = NVSS, F = FIRST, M = 2MASS, s = SDSS dr9, 6 = 6dFG;  QSO  = quasar, BL = BL Lac; v = variable in \wse\ bands 
(var\_flag $>$ 5 in at least one band, see Cutri et al. 2012 for additional details); rv = variable in the radio bands at 1.4 GHz. \\
Col. (9) Redshift: ? = unknown. \\
{ Note: The source WN 1514.8+3701 can be considered as a blazar-like source on the basis of the multifrequency investigation (see Section~\ref{sec:appendix} for more details)
even if its $\alpha_{325}^{1400}$ spectral index is not agreement with the type A and type B candidates.}
\label{tab:main}
\end{table*}

\begin{table}
\caption{Optical magnitudes of the USNO B1 catalog for the WENSS $\gamma$-ray blazar candidates}
\begin{tabular}{|lccccc|}
\hline
WENSS   &   B1     &   R1     &   B2     &   R2     &   I    \\ 
name    &   mag    &   mag    &   mag    &   mag    &   mag   \\ 
\hline
\noalign{\smallskip}
  WN 0140.0+8546 & 20.5 & 18.87 & 20.03 & 18.55 & 18.62\\
  WN 0245.0+5118 &  & 19.25 & 20.01 & 19.44 & 18.28\\
  WN 0303.9+4903 & 19.64 & 17.41 & 18.1 & 16.76 & 15.59\\
  WN 0327.5+6258 &  &  & 20.66 & 19.92 & 18.35\\
  WN 0349.1+5645 & 20.09 & 19.24 & 20.43 & 18.76 & 18.53\\
  WN 0557.5+3838 &  & 19.11 &  & 19.84 & 18.48\\
  WN 0640.0+6041 & 20.01 & 19.58 & 20.7 & 18.75 & 18.37\\
  WN 0720.7+2905 & 19.78 & 19.05 & 19.97 & 18.72 & \\
  WN 0839.4+6724 &  &  &  & 20.5 & 18.74\\
  WN 1012.4+5605 & 19.69 & 19.42 & 20.61 & 19.35 & \\
  WN 1222.0+8010 &  & 17.6 & 20.18 & 18.46 & 17.63\\
  WN 1500.0+4815 & 19.06 & 18.52 & 19.77 & 18.95 & 19.05\\
  WN 1501.2+5559 &  & 19.83 & 20.92 & 20.14 & \\
  WN 1514.8+3701 & 20.9 &  & 21.49 & 20.07 & 19.16\\
  WN 1826.8+3220 & 19.5 & 18.26 & 19.23 & 17.9 & 17.44\\
  WN 2106.1+3643 & 19.87 & 18.06 &  & 18.71 & \\
\noalign{\smallskip}
\hline
\end{tabular}\\
\label{tab:usno}
\end{table}

~\\
\acknowledgements
We thank the anonymous referee for useful comments that improve the presentation of our results.
F. Massaro thanks R. Morganti for her valuable suggestions and for her hospitality at ASTRON
together with E. Mahony and G. Heald for their discussion.
F. Massaro is also grateful to S. Digel, J. Grindlay, M. orienti, Howard Smith 
D. Thompson and M. Urry for their helpful discussions
as well as to M. Ajello, E. Ferrara and J. Ballet for their support.
The work is supported by the NASA grants NNX12AO97G.
R. D'Abrusco gratefully acknowledges the financial 
support of the US Virtual Astronomical Observatory, which is sponsored by the
National Science Foundation and the National Aeronautics and Space Administration.
The work by G. Tosti is supported by the ASI/INAF contract I/005/12/0.
The WENSS project was a collaboration between the Netherlands Foundation 
for Research in Astronomy and the Leiden Observatory. 
We acknowledge the WENSS team consisted of Ger de Bruyn, Yuan Tang, 
Roeland Rengelink, George Miley, Huub Rottgering, Malcolm Bremer, 
Martin Bremer, Wim Brouw, Ernst Raimond and David Fullagar 
for the extensive work aimed at producing the WENSS catalog.
Part of this work is based on archival data, software or on-line services provided by the ASI Science Data Center.
This research has made use of data obtained from the High Energy Astrophysics Science Archive
Research Center (HEASARC) provided by NASA's Goddard
Space Flight Center; the SIMBAD database operated at CDS,
Strasbourg, France; the NASA/IPAC Extragalactic Database
(NED) operated by the Jet Propulsion Laboratory, California
Institute of Technology, under contract with the National Aeronautics and Space Administration.
This research has made use of the VizieR catalogue access tool, CDS, Strasbourg, France.
Part of this work is based on the NVSS (NRAO VLA Sky Survey);
The National Radio Astronomy Observatory is operated by Associated Universities,
Inc., under contract with the National Science Foundation. 
This publication makes use of data products from the Two Micron All Sky Survey, which is a joint project of the University of 
Massachusetts and the Infrared Processing and Analysis Center/California Institute of Technology, funded by the National Aeronautics 
and Space Administration and the National Science Foundation.
This publication makes use of data products from the Wide-field Infrared Survey Explorer, 
which is a joint project of the University of California, Los Angeles, and 
the Jet Propulsion Laboratory/California Institute of Technology, 
funded by the National Aeronautics and Space Administration.
TOPCAT\footnote{\underline{http://www.star.bris.ac.uk/$\sim$mbt/topcat/}} 
\citep{taylor2005} for the preparation and manipulation of the tabular data and the images.
Funding for the SDSS and SDSS-II has been provided by the Alfred P. Sloan Foundation, 
the Participating Institutions, the National Science Foundation, the U.S. Department of Energy, 
the National Aeronautics and Space Administration, the Japanese Monbukagakusho, 
the Max Planck Society, and the Higher Education Funding Council for England. 
The SDSS Web Site is http://www.sdss.org/.
The SDSS is managed by the Astrophysical Research Consortium for the Participating Institutions. 
The Participating Institutions are the American Museum of Natural History, 
Astrophysical Institute Potsdam, University of Basel, University of Cambridge, 
Case Western Reserve University, University of Chicago, Drexel University, 
Fermilab, the Institute for Advanced Study, the Japan Participation Group, 
Johns Hopkins University, the Joint Institute for Nuclear Astrophysics, 
the Kavli Institute for Particle Astrophysics and Cosmology, the Korean Scientist Group, 
the Chinese Academy of Sciences (LAMOST), Los Alamos National Laboratory, 
the Max-Planck-Institute for Astronomy (MPIA), the Max-Planck-Institute for Astrophysics (MPA), 
New Mexico State University, Ohio State University, University of Pittsburgh, 
University of Portsmouth, Princeton University, the United States Naval Observatory, 
and the University of Washington.

{}

\appendix
\label{sec:appendix}
\subsection{Source details}
\label{sec:details}
Details for all the 32 UGSs analyzed are reported here.

\noindent
\underline{2FGL J0002.7+6220}: the unique WENSS source that lie within {\it search region} of radius $\theta_{95}$
does not appear to be a blazar-like source, since the radio spectrum looks steep
and the source, even if detected in \wse\ and in 2MASS, 
does not have the IR colors similar to those of the $\gamma$-ray blazars \citep[e.g.,][]{paper1,paper6}.

\noindent
\underline{2FGL J0039.1+4331}: the unique WENSS source WN 0036.7+4320 could be not considered a 
$\gamma$-ray blazar given its $\alpha_{325}^{1400}=$ 0.73 since this occurs only in few cases within the LGB sample 
(i.e., $\sim$3\%, see Figure~\ref{fig:hist_fermi}).

\noindent
\underline{2FGL J0158.6+8558}: the WENSS source WN 0144.4+8550 is not a good counterpart given its radio steep spectrum while
both the other two WENSS sources, that lie within the {\it search region}, 
have radio spectra consistent with the $\gamma$-ray blazar population.
Unfortunately, for these two WENSS sources no additional multifrequency information 
is available in literature to clearly distinguish the best candidate counterpart.
In this case we favor the WENSS source WN 0144.4+8546 with respect to WN 0140.0+8546 
as candidate counterpart of 2FGL J0158.6+8558, since it lies closer to the $\gamma$-ray centroid position.

\noindent
\underline{2FGL J0248.5+5131}: the radio source WN 0245.0+5118 is the preferred counterpart,
since it has a flat radio spectrum (i.e.,  $\alpha_{325}^{1400}=$0.36) 
and it is detected in both the \wse\ and the 2MASS catalogs,
as occurs for typical $\gamma$-ray emitting blazars \citep[e.g.][]{paper6}.

\noindent
\underline{2FGL J0307.4+4915}: the unique WENSS source is a $\gamma$-ray emitting blazar candidate of type A, 
possible counterpart, given its flat radio spectrum, its compact radio structure (see Figure~\ref{fig:radio})
and its IR \wse\ colors (see Figure~\ref{fig:plane1}).

\noindent
\underline{2FGL J0332.1+6309}: this UGS appear to be associated with WN 0327.5+6258, $\gamma$-ray blazar candidate of type A,
as occurs for the previous case of 2FGL J0307.4+4915.

\noindent
\underline{2FGL J0336.0+7504}: the steep radio spectrum of the WENSS source WN 0330.3+7500 within the 
{\it search region} looks different from that of the LGB sample, 
thus WN 0330.3+7500 does not appear to be a $\gamma$-ray blazar candidate.

\noindent
\underline{2FGL J0353.2+5653}: the source WN 0349.1+5645 
clearly show many features of the $\gamma$-ray blazar candidates with respect to 
the other two WENSS sources present within the $\theta_{95}$ {\it search region}. 
It has the typical flat spectrum as the sources in the LGB sample, and moreover,  
it is also variable in the radio band according to the archival 1.4 GHz observations \citep{white92,condon98}.

\noindent
\underline{2FGL J0600.9+3839}: WN 0557.5+3838 is a $\gamma$-ray blazar candidate of type B, being also detected by \wse.

\noindent
\underline{2FGL J0644.6+6034}: WN 0640.0+6041 is similar to the previous case of 2FGL J0600.9+3839.

\noindent
\underline{2FGL J0723.9+2901}: WN 0720.7+2905 is similar to the previous case of 2FGL J0644.6+6034, however, given 
its flatter radio spectrum it is a type A candidate.

\noindent
\underline{2FGL J0843.6+6715}: WN 0839.4+6724 is most probably the counterpart of the UGS 2FGL J0843.6+6715 source,
being a $\gamma$-ray blazar candidate of type A, while
WN 0839.7+6728 even if detected by \wse\ has IR colors very different from those of the $\gamma$-ray blazars \citep{paper6}.

\noindent
\underline{2FGL J0844.9+6214}: WN 0839.8+6230, given its steep radio spectrum 
is unlikely to be a $\gamma$-ray blazar candidate associable to 2FGL J0844.9+6214.

\noindent
\underline{2FGL J0900.9+6736}: WN 0856.1+6754 is similar to the previous case of 2FGL J0723.9+2901 even if the spectral index 
is only marginally inconsistent with those of the LGB sample (see Figure~\ref{fig:hist_fermi}).

\noindent
\underline{2FGL J1013.6+3434}: WN 1010.4+3445 is similar to the case of 2FGL J0900.9+673, being also detected in the SDSS with 
the typical blue optical colors of a blazar-like source \citep{massaro12}.

\noindent
\underline{2FGL J1016.1+5600}: within all the WENSS sources 
that lie within the $\theta_{95}$ positional uncertainty region of this UGS, 
WN 1012.4+5605 shows features similar to those of $\gamma$-ray blazar candidates, 
having a flat radio spectrum and a \wse\ counterpart with the IR colors consistent with those
of the \wse\ Gamma-ray strip \citep[e.g.][see also Figure~\ref{fig:plane1}]{paper3,paper6}. 
The source WN 1014.2+5616 even if with a flat radio spectrum was not considered 
as a $\gamma$-ray blazar candidate on the basis of its \wse\ IR colors, too unusual for blazars.
On the other hand, also WN 1013.6+5616, can be considered as a $\gamma$-ray blazar candidate of type B,
but less multifrequency observations and the lack of \wse\ counterpart do not favor it.
It is worth noting that the source WN 1013.3+5629 (alias NVSS J101640+561445) it is a clear FR\,II radio galaxy \citep{fanaroff74},
which low frequency emission is dominated by its lobes and with a typical steep radio spectrum
very different from those of the other potential counterparts.

\noindent
\underline{2FGL J1223.3+7954}: WN 1222.0+8010 is similar to case of 2FGL J0332.1+6309.

\noindent
\underline{2FGL J1502.1+5548}: in this UGS there are three candidate counterparts, two of type B and one of type A.
The latter is the favored counterpart being of class A, even if also WN 1501.2+5559 a $\gamma$-ray blazar candidate of type B
has multifrequency observations that indicate it as a potential blazar-like source.

\begin{figure}[!t] 
           \includegraphics[height=9.6cm,width=6.6cm,angle=-90]{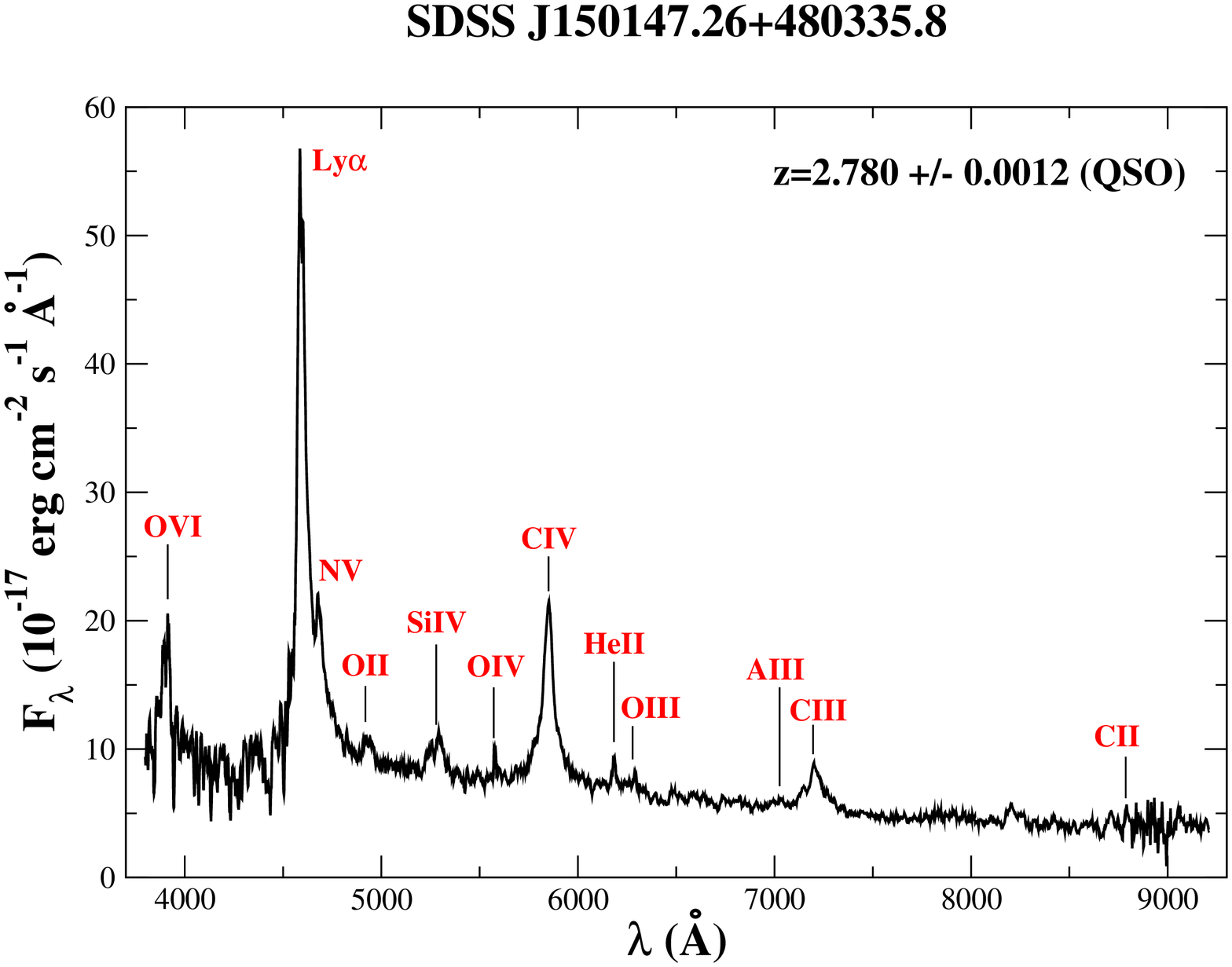}
           \includegraphics[height=9.6cm,width=6.6cm,angle=-90]{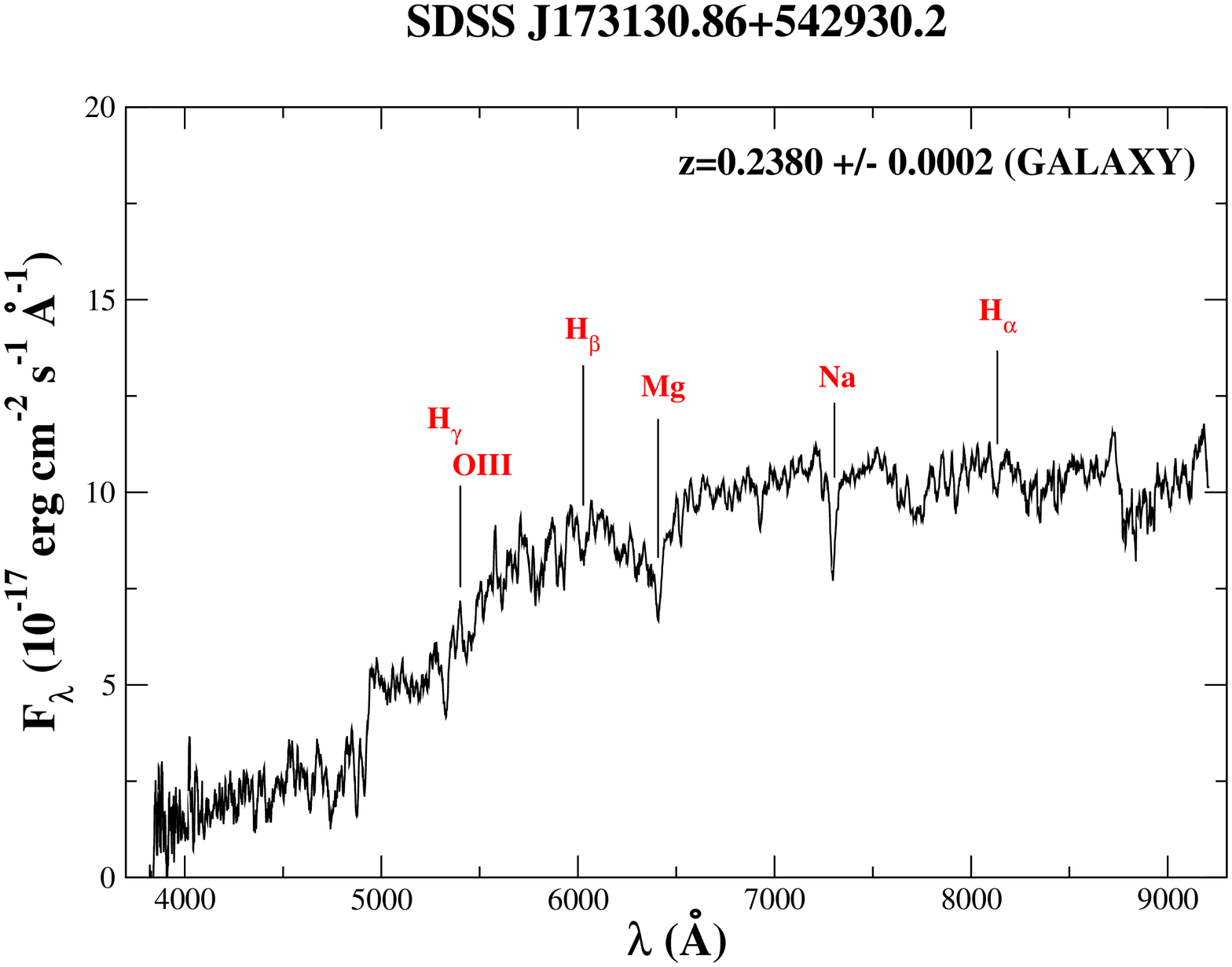}
           \caption{Left panel) The archival SDSS spectroscopic observation 
                        of the $\gamma$-ray blazars candidate of type A: WN 1500.0+4815
                        associated with the \fer\ source 2FGLJ1502.4+4804. The prominent Ly$\alpha$ emission line 
                        together with several Carbon emission lines allows a clear redshift estimate of 2.78.
                        It is the 7$^{th}$ most distant $\gamma$-ray source in the 2LAC and 2FGL catalogs.
                        Right panel) The archival SDSS optical spectrum of the radio steep spectrum source WN 1730.4+5431, typical of a galaxy at z=0.238.}
          \label{fig:sdss_spectrum}
          \end{figure}

\noindent
\underline{2FGL J1502.4+4804}: the radio source WN 1500.0+4815 is
a $\gamma$-ray blazar candidate of type A candidate counterpart of thsi UGS. 
It appears similar to a BZQ source at high redshift, 
with its optical SDSS spectrum, shown in Figure~\ref{fig:sdss_spectrum} (left panel) \citep[e.g.,][]{adelman08,paris12}. 
This association is the 7$^{th}$ most distant $\gamma$-ray source known 
according to the 2LAC and the 2FGL catalogs\citep[][respectively]{ackermann11a,nolan12}.
It has a clear radio compact structure (see Figure~\ref{fig:radio})
and its IR colors are consistent with those of the $\gamma$-ray blazars in the [3.4]-[4.6]-[12] $\mu$m diagram 
as shown in Figure~\ref{fig:plane1}.

\noindent
\underline{2FGL J1517.2+3645}: 
among the WENSS sources that lie in the {\it search region} of 2FGL J1517.2+3645, 
only WN 1514.8+3701 can be considered a gamma-ray blazar candidates being variable at 1.4 GHz radio frequencies 
as for the case of 2FGL J0353.2+5653, even if with a steep radio spectrum (see Section~\ref{sec:ugs}).

\noindent
\underline{2FGL J1614.8+4703}: the flat spectrum of WN 1612.6+4714 makes it the most probable counterpart 
of this UGS with respect to the other nearby WENSS sources, according to our association criteria (see Section~\ref{sec:ugs}).

\noindent
\underline{2FGL J1623.2+4328}: similar to the cases of 2FGL J0039.1+4331 we did not consider 
its WENSS source a potential blazar-like counterpart.
Even if WN 1621.7+4341 is detected by \wse\ at both 3.4$\mu$m and 4.6$\mu$m as occurs for the 
large majority of the \bzcat\ sources \citep[e.g.,][]{paper2,paper6}, its radio spectrum is 
steeper that those of the LGB sources (see Figure~\ref{sec:index}).

\noindent
\underline{2FGL J1730.8+5427}: this UGS does not present any $\gamma$-ray blazar candidates associable.
Both the WENSS sources in its {\it search region} have steep radio spectra.
Moreover the radio source WN 1730.4+5431 has also the SDSS optical spectrum typical of an elliptical galaxy
with a redshift estimate of 0.238 \citep[][see also Figure~\ref{fig:sdss_spectrum}, right panel]{paris12}.

\noindent
\underline{2FGL J1738.9+8716}: WN 1758.4+8718 could be considered a $\gamma$-ray blazar 
candidate on the basis of its flat radio spectrum; however it has been discarded
on the basis of its IR colors are not those typical of the $\gamma$-ray emitting blazars \citep{paper4,paper6}.

\noindent
\underline{2FGL J1748.8+3418}: similar to the case of 2FGL J0644.6+6034, 
the WENSS source WN 1747.6+3424 appear to be a $\gamma$-ray blazar candidate of type B,
potential counterpart of this UGS.

\noindent
\underline{2FGL J1828.7+3231}: the flat radio spectrum of WN 1826.8+3220 
and its \wse\ detection suggest that it could be $\gamma$-ray blazar candidate of type A (see also Figure~\ref{fig:plane1}).

\noindent
\underline{2FGL J2041.2+4735}: the steep spectrum of WN 2039.6+4728 
does not appear to be similar to those of the sources in the LGB sample (see Section~\ref{sec:index}).

\noindent
\underline{2FGL J2107.8+3652}: the inverted spectrum of WN 2106.1+3643, 
coupled with its \wse\ detection points it as $\gamma$-ray blazar candidate of type A (see also Figure~\ref{fig:plane1}).

\noindent
\underline{2FGL J2110.3+3822}: the marginally steep spectrum of WN 2108.3+3804 does not allow to claim it as 
a $\gamma$-ray blazar candidate according to our criteria, despite its \wse\ and 2MASS IR detections.

\noindent
\underline{2FGL J2114.1+5440}: WN 2113.6+5429 is a similar case to 2FGL J2041.2+4735 with a flatter spectrum.

\noindent
\underline{2FGL J2133.9+6645}: the flat spectrum of the single radio source WN 2133.6+6625 
lying within the {\it search region} indicates this source as a $\gamma$-ray blazar candidate of type A.

\noindent
\underline{2FGL J2231.0+6512}: similar to the previous case of 2FGL J2133.9+6645, 
we favor the flat spectrum radio source WN 2229.1+6451
as potential counterpart of this UGS, while the other WENSS source WN 2230.2+6450 
has a steep radio spectrum not seen in other sources of the LGB sample.

\end{document}